\documentclass[sigconf, screen]{acmart}

\AtBeginDocument{%
  \providecommand\BibTeX{{%
    \normalfont B\kern-0.5em{\scshape i\kern-0.25em b}\kern-0.8em\TeX}}}

\copyrightyear{2026}
\acmYear{2026}
\setcopyright{cc}
\setcctype{by}
\acmConference[CHI '26]{Proceedings of the 2026 CHI Conference on Human Factors in Computing Systems}{April 13--17, 2026}{Barcelona, Spain}
\acmBooktitle{Proceedings of the 2026 CHI Conference on Human Factors in Computing Systems (CHI '26), April 13--17, 2026, Barcelona, Spain}
\acmPrice{}
\acmDOI{10.1145/3772318.3790585}
\acmISBN{979-8-4007-2278-3/2026/04}

\usepackage{subcaption} % captions for subfigures and the like

\usepackage{url}
\makeatletter
\g@addto@macro{\UrlBreaks}{\UrlOrds}
\makeatother

\usepackage{multirow} %tabular cells spanning multiple rows
\usepackage{color} %colour management
\usepackage{enumitem} %better  enumerate, itemize, and description

\usepackage{xspace}
\usepackage{xcolor, colortbl}% http://ctan.org/pkg/xcolor

\usepackage{arydshln}

\newcommand{\m}{\textit{M=}}

\newcommand{\sd}{\textit{SD=}}

\newcommand{\F}[3]{$F({#1},{#2})={#3}$}

\newcommand{\p}{\textit{p=}}

\newcommand{\rankbiserial}[1]{$r_{rb} = #1$}

\newcommand{\pminor}{\textit{p$<$}}

\newcommand{\padj}{\textit{p$_{adj}$=}}

\newcommand{\Scenario}{road user role\xspace}

\def\plainkeywords{External communication; Autonomous vehicles; Chicken Game; Pedestrian Behavior; eHMI.}

\begin{document}

%[short-title]
\title[eHMI for All]{eHMI for All - Investigating the Effect of External Communication of Automated Vehicles on Pedestrians, Manual Drivers, and Cyclists in Virtual Reality}

\author{Mark Colley}
\email{m.colley@ucl.ac.uk}
\orcid{0000-0001-5207-5029}
\affiliation{%
  \institution{Institute of Media Informatics, Ulm University}
  \city{Ulm}
  \country{Germany}
}
\affiliation{%
  \institution{UCL Interaction Centre}
  \city{London}
  \country{United Kingdom}
}

\author{Simon Kopp}
\email{simon.kopp@uni-ulm.de}
\orcid{0000-0003-2071-2374}
\affiliation{%
  \institution{Institute of Media Informatics, Ulm University}
  \city{Ulm}
  \country{Germany}
}

\author{Debargha Dey}
\email{d.dey@tue.nl}
\orcid{0000-0001-9266-0126}
\affiliation{%
  \institution{Eindhoven University of Technology}
  \city{Eindhoven}
  \country{The Netherlands}
 }

\author{Pascal Jansen}
\email{pascal.jansen@uni-ulm.de}
\orcid{0000-0002-9335-5462}
\affiliation{%
  \institution{Institute of Media Informatics, Ulm University}
  \city{Ulm}
  \country{Germany}
}

\author{Enrico Rukzio}
\email{enrico.rukzio@uni-ulm.de}
\orcid{0000-0002-4213-2226}
\affiliation{%
  \institution{Institute of Media Informatics, Ulm University}
  \city{Ulm}
  \country{Germany}
}

% 150 words or less
% What is the large scope and problem space? Why should we care? Motivation

% What is the specific problem addressed? Problem

% Why is the problem Important? Why was this work carried out?

% What have you done? Solution

% What did you find out? What are the concrete results?

% What are the implications on a larger scale? How does it change the bigger picture?

\renewcommand{\shortauthors}{Colley et al.}

% Do not change the page size or page settings.
\begin{abstract}
With automated vehicles (AVs), the absence of a human operator could necessitate external Human-Machine Interfaces (eHMIs) to communicate with other road users. Existing research primarily focuses on pedestrian-AV interactions, with limited attention given to other road users, such as cyclists and drivers of manually driven vehicles. So far, no studies have compared the effects of eHMIs across these three road user roles. 
Therefore, we conducted a within-subjects virtual reality experiment (N=40), evaluating the subjective and objective impact of an eHMI communicating the AV's intention to pedestrians, cyclists, and drivers under various levels of distraction (no distraction, visual noise, interference). eHMIs positively influenced safety perceptions, trust, perceived usefulness, and mental demand across all roles. While distraction and road user roles showed significant main effects, interaction effects were only observed in perceived usability. Thus, a unified eHMI design is effective, facilitating the standardization and broader adoption of eHMIs in diverse traffic.
\end{abstract}




\maketitle

\section{Introduction}
%Motivation
Automated vehicles (AVs) with an SAE (Society of Automobile Engineers) level~\cite{taxonomy2014definitions} of 4 or 5 will fundamentally transform traffic~\cite{fagnant2015preparing} and interactions within traffic environments~\cite{colley2020adesign, dey2020taming, locken2019should}.
As no human operator needs to be present to communicate with other road users in situations with uncertainty (e.g., regarding right of way), AVs must be equipped with the ability to communicate with other road users in ambiguous situations. This is done through external Human-Machine Interfaces, or short, eHMIs. Previous work proposed numerous eHMIs. For example, displays on windshields~\cite{dey2020distancedependent}, LED strips~\cite{RefWorks:doc:5cf7ad8de4b06bba938e0112, RefWorks:doc:5cf8aa47e4b006bc06a90b0a, 10.1145/3706598.3714187}, anthropomorphic or animal-related communication~\cite{10.1145/3765766.3765767}, projections~\cite{ackermann2019experimental, nguyen2019designing}, and implicit communication such as movement patterns~\cite{zimmermann2017first}. External devices like smartphones~\cite{hollander2020smombies} and enhanced infrastructure have also been considered~\cite{RefWorks:doc:5cd92775e4b0487541989799}. Furthermore, studies have already explored different eHMI concepts and designs, like text-based designs~\cite{chang2018video} that communicate vehicle status~\cite{faas2020longitudinal} or intention~\cite{dietrich2020automated}. Studies have shown mainly positive effects of eHMIs on trust, clarity, hedonic qualities, and pedestrian crossing behaviors.

%Problem
However, most of them only consider communication between one pedestrian and one AV \cite{10.1145/3334480.3382865, colley2020scalability}. Only relatively little work has been conducted with cyclists~\cite{10.1145/3313831.3376884, FERENCHAK202331} or drivers of ordinary, manually driven vehicles (MDVs)~\cite{10.1145/3546711, RETTENMAIER2020175, 8814082}. No work so far has evaluated the effects of eHMIs on all three road users comparatively. Furthermore, it is important to ascertain if a chosen eHMI can function effectively across different kinds of road users to prevent the situation that an AV is equipped with multiple kinds of eHMIs, each catering to a different road user. This can lead to a busy interface, which can, in turn, cause information overload, leading to detrimental effects in the user interface. It is therefore crucial to explore if a holistic approach to eHMI design can work- i.e., whether a unified eHMI is able to perform adequately across different kinds of road users. In this work, we address this gap by employing one unified scene and one eHMI design integrating simulators for MDVs, pedestrians, and cyclists. 
 
Additionally, more comprehensive studies that scale up eHMI experiments are required for the results and subsequent insights to be more externally valid. \citet{colley2019better} have identified over 38 factors influencing pedestrian crossing decisions, including physical context, dynamic factors, traffic characteristics, social factors, demographics, abilities, and personal characteristics. While some of these factors are dependent on personal, cultural, or national contexts (e.g., demographics, social norms, and law compliance), others relate directly to the external communication of AVs. Therefore, in this work, we also study the influence of distractions (visual noise) or interference (the trajectories of the AV and other traffic intersect, thereby causing an interaction)~\cite{8667866, Markkula2020}.

%Solution
We conducted a within-subjects virtual reality (VR) study with $N = 40$ participants to evaluate the effects of eHMIs (no eHMI and an intention-based eHMI) on the participants being in the role of three different road users (pedestrian, driver, cyclist) in scenarios with different levels of distraction (no distraction, visual noise, and interference).

In this study with multiple AVs, MDVs, and other pedestrians, eHMIs had positive effects. In these scenarios, the participants generally felt safer, trusted the AVs more, assessed the system as more useful, and felt less mentally demanded. While we also found that distraction and the road user role have significant main effects, we did not find interaction effects except for usability. As a result, we posit that one unified eHMI concept is applicable for diverse road user roles, potentially significantly improving the standardization and adoption of eHMIs.
%These tendencies showed that confusion could occur in scenarios with more than one road user and that there is also a need for more studies on communication between AVs and manual drivers or AVs and cyclists.

\textit{Contribution Statement:} This research offers insights into the effects of external communication of AVs for pedestrians, cyclists, and manual drivers. Results from our VR study with $N = 40$ participants suggest that eHMIs are effective across different road user roles. This highlights the potential for a holistic approach to a unified eHMI design that caters to multiple types of road users.

\section{Related Work}
We ground our work in existing research on pedestrian-vehicle interaction, specifically, crossing behaviors in traffic and the space and impact of eHMIs of AVs on traffic interactions.

\subsection{External Communication of Automated Vehicles}
Current traffic interactions often rely on gestures and eye contact to resolve ambiguities, for example, regarding the right of way~\cite{rasouli2017understanding}. Despite the infrequent necessity for explicit communication~\cite{lee2021road}, eHMIs have been proposed to facilitate communication between AVs and vulnerable road users like pedestrians or cyclists~\cite{hollander2021taxonomy}. Prior work has categorized external communication strategies by modality, message type, and location on the vehicle~\cite{colley2020adesign, colley2020design, dey2020taming}. \citet{colley2020adesign} identified eight message types: Instruction, Command, Advisory, Answer, Historical, Predictive, Question, and Affective. Communication locations include various parts of the vehicle, personal devices, or infrastructure, such as sidewalks, with specific focus areas like the windshield or bumper highlighted as key interaction points~\cite{dey2019gaze}.

Key considerations for effective eHMI deployment include the communication relationship dynamics (ranging from one-to-one to many-to-many), ambient noise levels, and the specific road user involved (e.g., pedestrian, cyclist)~\cite{colley2020adesign}. Studies have explored eHMI effectiveness across diverse groups, including children~\cite{deb2019comparison, RefWorks:doc:5cf7c9e4e4b03d2faef34312}, individuals with vision or mobility impairments~\cite{colley2020towards, asha2021co}, and general pedestrians~\cite{ackermans2020effects, dey2018interface, locken2019should, Dey2024multimodal}, as well as cyclists~\cite{hou2020autonomous}. Mostly positive outcomes of eHMIs have been reported. For instance, \citet{dey2020distancedependent} demonstrated that distance-dependent information could significantly improve pedestrians' understanding of AV intentions and willingness to cross safely. In contrast, \citet{colley2020towards} highlighted the preference among visually impaired individuals for clear, speech-based communications over other forms.

Despite these advancements, some challenges remain, such as ensuring children properly interpret and use eHMIs~\cite{deb2019comparison}. Moreover, concerns about overtrust~\cite{hollander2019overtrust}, the scalability of eHMIs~\cite{colley2020scalability, colley2023scalability}, and the exploration of eHMIs' social implications~\cite{sadeghian2020exploration, colley2021investigating, lanzer2020designing, sahin2021workshop} indicate ongoing research needs.

This study extends the exploration of eHMIs by implementing a simulated Slow-Pulsing Light Band (SPLB) eHMI, chosen among many proposed eHMI designs in the state of the art for its technological feasibility and language independence~\cite{LEE2022270, dey2020color}. This eHMI addresses a broader demographic, including children and non-native speakers. The light band of the SPLB eHMI is enhanced by extending it beyond just the front bumper to all sides of the AV.

\subsection{Simulators in eHMI Research}
Three distinct simulators are necessary to compare the effect of road user type (pedestrians, drivers of ordinary cars, and cyclists) combined with eHMIs. This is crucial because each simulator addresses different simulation requirements for accurately displaying the scene and enabling haptic interaction with specific props like a bicycle or steering wheel.

Regarding pedestrian simulators, most prior work employed VR headsets to simulate pedestrian environments, which provided high immersion by allowing participants to move freely in a controlled space and often included auditory enhancements to simulate real-world sounds~\cite{10.1145/3581961.3609875, 10.1145/3594806.3594827, 10.1145/3544548.3581303, 10.1145/3581961.3610373}. Others have used a cave automatic virtual environment (CAVE), which is typically a projection-based video theater in a larger room. CAVE-based simulators offer an immersive virtual reality environment where projectors are directed to between three and six of the walls of a room-sized cube~\cite{Kaleefathullah2020, Yang2023, LEE2022270}, although these are resource-intensive to set up and operate. Fewer studies used a monitor-based setup accessed through a webpage, offering lower immersion but noted for its accessibility and minimal spatial requirements~\cite{10.1145/3596248}. Some studies have already used actual vehicles equipped with eHMIs, providing the highest level of realism but at a higher cost and logistical complexity~\cite{10.1145/3581961.3609883, 10.1145/3544549.3585629, 10.1145/3580585.3606460, Dey2020_eHMI_gaze}.

Simulators for manual drivers have been used less frequently. One employed a monitor setup with a steering wheel and pedals, which provided a controlled environment for safe behavior analysis but with lower immersion~\cite{10.1145/3580585.3607155}. VR was also used for evaluating effects on manual drivers at intersections~\cite{10.1145/3546711}. Another study used a real-life scenario with a hidden driver in a vehicle equipped with an eHMI, offering a high degree of realism and naturalistic observation opportunities~\cite{10.1145/3580585.3607173}.

Regarding cyclist simulators, a VR headset with a stationary bike setup with sensors to measure steering angle and speed, offering a safe and controlled environment with high immersion, was often used~\cite{10.1145/3581961.3609849, hou2020autonomous}. One work already involved a real-life interaction between a bicycle and various vehicles, providing the most realistic interaction dynamics but requiring extensive setup and being affected by external conditions such as weather~\cite{10.1145/3580585.3607161}.

Overall, the choice of simulator technology varied based on research goals. In previous works, VR setups provided a good balance between realism and control, monitor-based setups were noted for their accessibility and ease of use, and real-life scenarios offered the most immersive experience but with higher costs and complexity. Based on these considerations, we opted for VR simulators for all three road user types.

\section{Experiment Setup}
We used Unity~\cite{unitygameengine} (version 2021.3) to set up the simulator and evaluate the effects of eHMIs on different road user roles: pedestrians, manual drivers, and cyclists.

\subsection{Pedestrian, Manual Driver, and Cyclist Simulator}

\begin{figure*}[ht]
    \centering
    \begin{subfigure}[b]{0.32\linewidth}
        \centering
        \includegraphics[width=\linewidth]{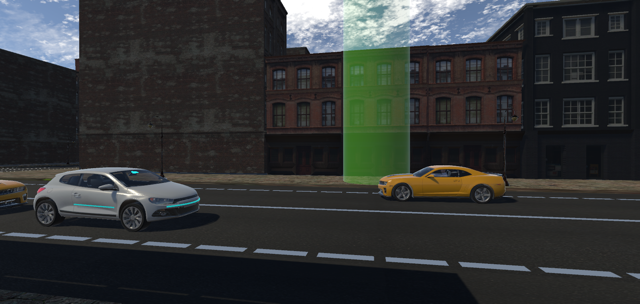}
        \caption{The pedestrian's point of view.}
        \label{fig:pedestrian-pov}
    \end{subfigure}
    \hfill
    \begin{subfigure}[b]{0.32\linewidth}
        \centering
        \includegraphics[width=\linewidth]{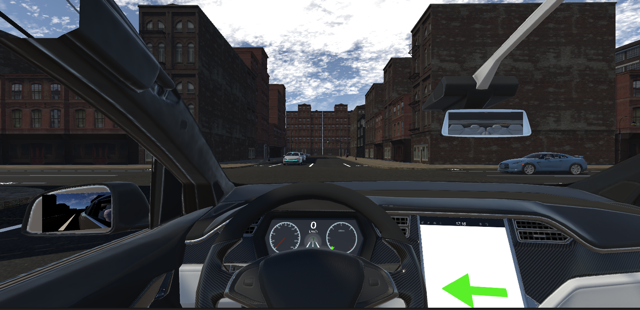}
        \caption{The driver's point of view.}
        \label{fig:driver-pov}
    \end{subfigure}
    \hfill
    \begin{subfigure}[b]{0.32\linewidth}
        \centering
        \includegraphics[width=\linewidth]{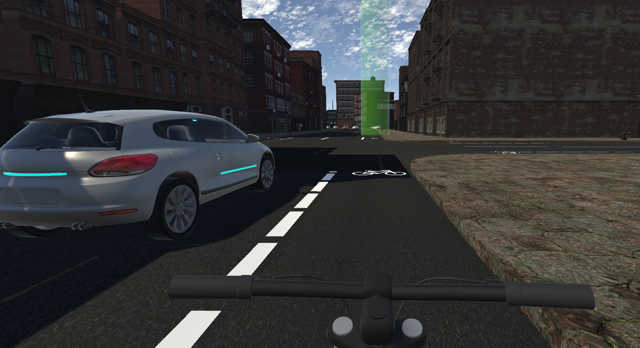}
        \caption{The bicyclist's point of view.}
        \label{fig:bicycle-povs}
    \end{subfigure}
    \caption{The three points of view for the pedestrian, driver, and cyclist scenarios.}
    \label{fig:povs}
    \Description{The three points of view for the pedestrian, driver, and cyclist scenarios.}
\end{figure*}

Participants in the role of pedestrians could walk freely through scenarios, drivers required a gaming steering wheel and pedals (with automatic transmission) for navigation, and cyclists used a bicycle setup attached to a speedometer, with an additional wind machine for realistic feedback.

\begin{figure*}[ht]
    \centering
    \begin{subfigure}[b]{0.32\linewidth}
        \centering
        \includegraphics[width=0.85\linewidth]{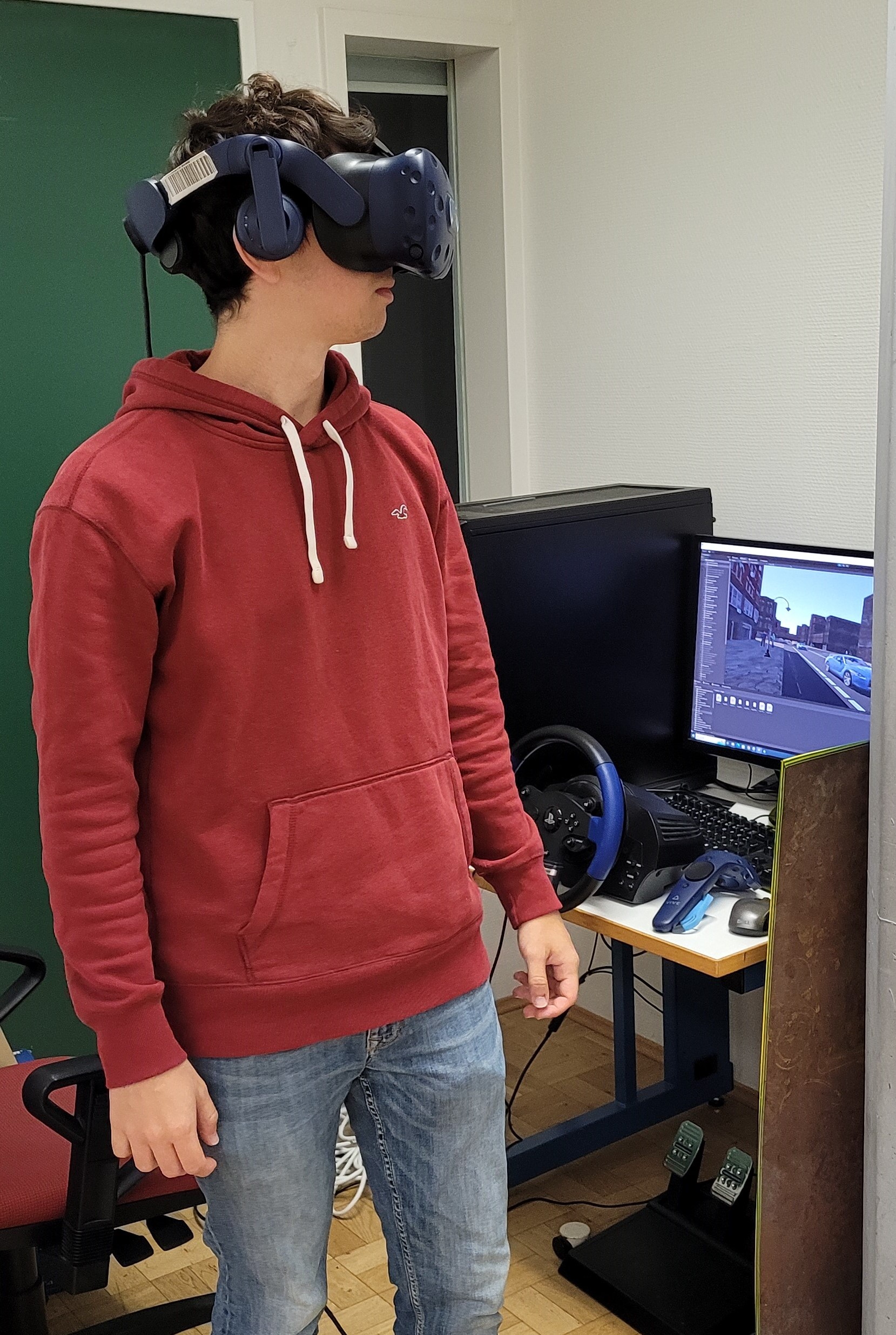}
        \caption{The pedestrian simulator. The space to walk (not visible) was about 7x2m.\newline}
        \label{fig:pedestrian-simulator}
    \end{subfigure}
    \hfill
    \begin{subfigure}[b]{0.32\linewidth}
        \centering
        \includegraphics[width=\linewidth]{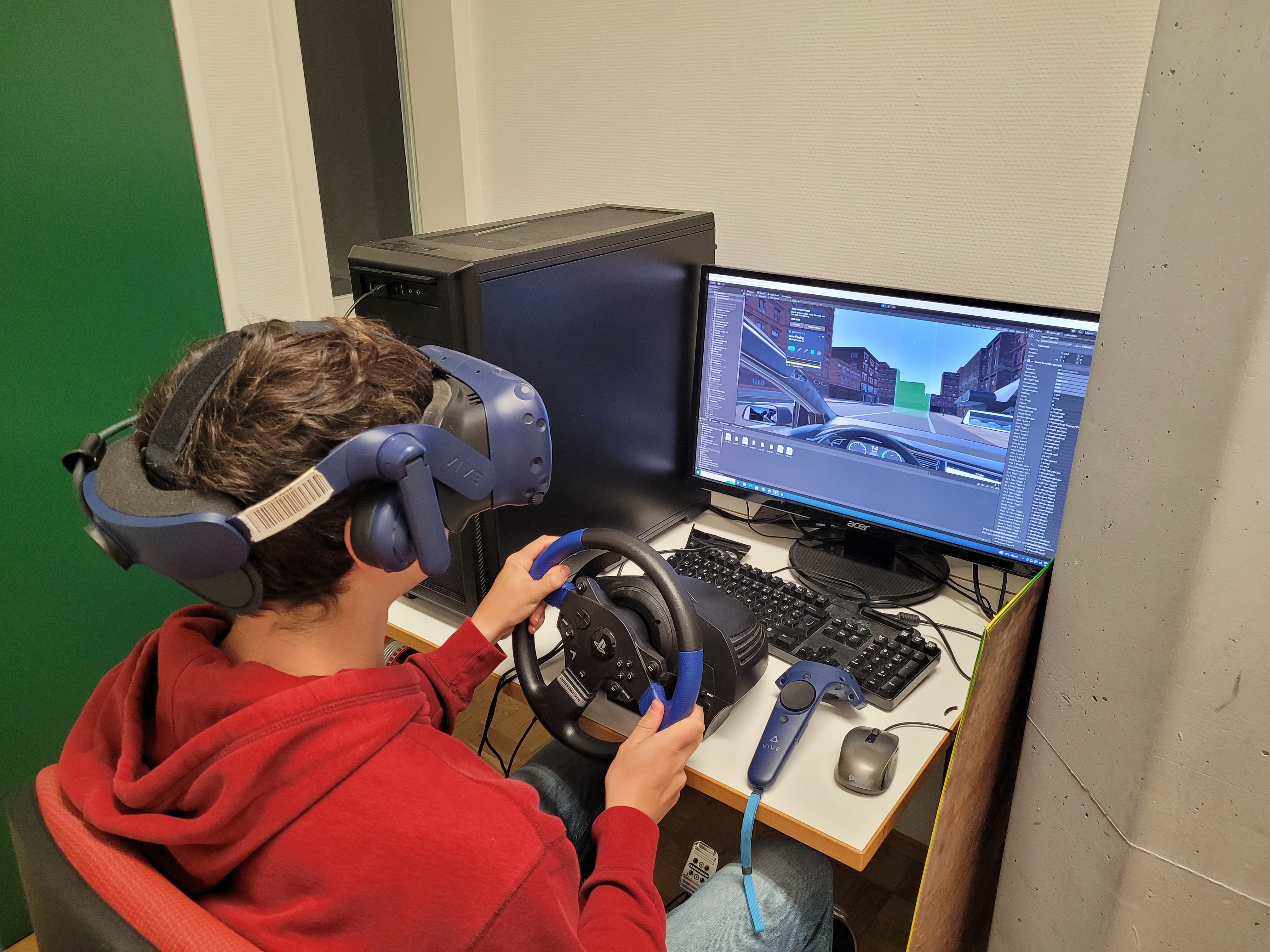}
        \caption{The driver simulator.\newline\newline}
        \label{fig:driver-simulator}
    \end{subfigure}
    \hfill
    \begin{subfigure}[b]{0.32\linewidth}
        \centering
        \includegraphics[width=0.85\linewidth]{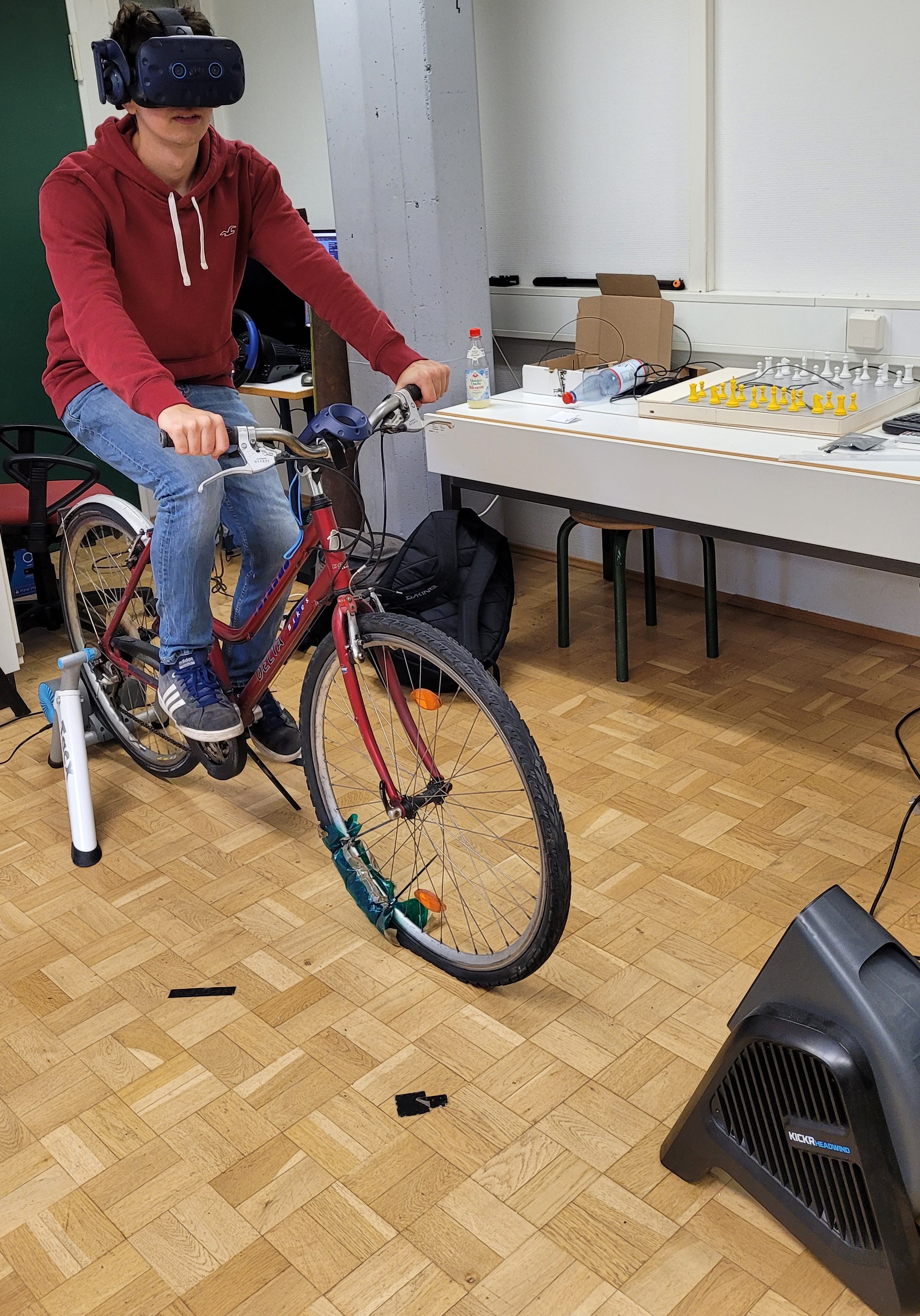}
        \caption{The bicycle simulator. The bicycle is connected to a speedometer (on the bike's rear wheel) and a wind machine (bottom right).}
        \label{fig:bicycle-simulator}
    \end{subfigure}
    \caption{The three simulator setups for the project are a pedestrian, a driver, and a cyclist. All setups are connected to the same computer and done with the same VR headset.}
    \label{fig:simulators-for-vr-study}
    \Description{The three simulator setups for the project are a pedestrian, a driver, and a cyclist. All setups are connected to the same computer and done with the same VR headset - an HTC Vive Pro Eye.}
\end{figure*}

\autoref{fig:simulators-for-vr-study} shows the three setups. \autoref{fig:povs} shows the points of view for the three road user roles.
An HTC VIVE Pro Eye VR headset was used.
For the manual driver simulator (see \autoref{fig:driver-simulator}), we used the steering wheel \href{https://www.thrustmaster.com/de-de/products/t150-force-feedback/}{Thrustmaster 150 Pro} with a pedal. The vehicle in Unity was $5 m$ in length, $2.05 m$ in width, and a height of $1.7 m$. This approximately represents a typical medium-sized C-segment vehicle. The bicycle simulator (see \autoref{fig:bicycle-simulator}) includes a bike and two \href{https://www.garmin.com/en-US/p/10997}{Ant+ USB Dongles}. These dongles communicate with the \href{https://www.garmin.com/de-DE/p/690890}{Tacx bicycle speedometer} and the \href{https://eu.wahoofitness.com/devices/indoor-cycling/accessories/kickr-headwind-buy-eu}{Wahoo Kickr wind machine}. The speedometer tracks how fast the person is pedaling and adjusts the speed in the virtual world. Internal tests showed that the wind from the wind machine increased immersion and decreased induced motion sickness.

\subsection{External Communication Concept}

\begin{figure*}[ht]
    \centering
    \begin{subfigure}[b]{0.38\linewidth}
        \centering
        \includegraphics[width=\linewidth]{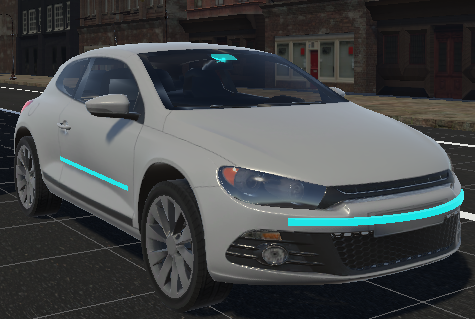}
        \caption{Front-view.}
        \label{fig:car-ehmi}
    \end{subfigure}
    \hfill
    \begin{subfigure}[b]{0.42\linewidth}
        \centering
        \includegraphics[width=\linewidth]{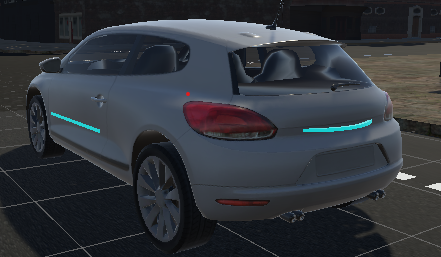}
        \caption{Rear-view.}
        \label{fig:car-ehmi-back}
    \end{subfigure}
    \caption{The white AV with the implemented intention-based SPLB. It blinks when it fully stops. A light band on the AV's side and rear supports omnidirectional communication. The light signal on the windshield around the rearview mirror indicates that it is an AV.}
    \label{fig:av-car}
    \Description{The white AV is based on a VW Scirocco with the implemented intention-based SPLB. It blinks when it fully stops. A light band on the AV's side and rear supports omnidirectional communication. The light signal on the windshield around the rearview mirror indicates that it is an AV.}
\end{figure*}

The Slow-Pulsing Light Band (SPLB) eHMI was designed to communicate the AV's intention of stopping, for example, for pedestrians wanting to cross the street, for cyclists wanting to continue on a street while the AV wants to turn, or for an intersection with other manual vehicles  (see \autoref{fig:av-car}). The choice of the SPLB is rooted in its popularity in eHMI research due to its relative simplicity, ease of implementation, and the possibility to communicate an AV's yielding intent without it appearing like an instructional or advisory message~\cite{ackermann2019experimental, Cefkin2019, de2019external, dey2018interface, Faas2019, habibovic2018communicating, Hamm2018, hensch2019should, petzoldt2018potential}. We adapted the light band design by integrating insights from prior research~\cite{dey2020color, Faas2019, hensch2019should, Hensch2019_steady_flashing_sweeping, kreissig_enlightening_2021}, which show that a uniform pattern, such as a slow-pulsing animation in a cyan color, is effective in showing yielding intention to, and extended it to all sides of the vehicle. In our implementation, it is attached to all four sides of the AV (see \autoref{fig:car-ehmi-back}) and is turned on and blinks when the AV intends to yield for the other road user. 

\subsection{Scenarios}
First, we describe the general (baseline) scenario, followed by two \textit{distraction} conditions: visual noise and interference. We define ``noise'' in a scene as situations where there are other traffic participants in the environment, even though their trajectories do not intersect that of the ego participant in the experiment. Scenes with ``interference'', on the other hand, characterize situations where the trajectory of another road user directly intersects with (and therefore interferes with) the intended path of the ego participant, causing an ``interaction'' as defined by \citet{Markkula2020}. As opposed to noise and interference, the baseline scenario does not include any other road user besides the participant and the AV. For brevity, only figures representing the scenarios with noise or interference are shown. However, in the control condition, the scenarios remain identical except for the noise or interference. We provide both a schematic and an implementation view. 

We considered distraction an important factor --- and consequently an independent variable --- in our study because naturalistic traffic interactions rarely unfold as one-on-one encounters in a clean, controlled setting. Yet, much of the existing eHMI literature continues to evaluate AV/ road-user communication in highly sanitized dyadic scenarios, limiting ecological validity and scalability of findings~\cite{dey2020taming}. Prior work has shown that visual and cognitive distraction influence traffic interactions~\cite{bazilinskyy2020examining, Tian2022_distraction, lanzer_interaction_2023}, i.e., complexity and the presence of additional road users can influence gaze allocation, situational assessment, and road-crossing decisions. When an eHMI is activated in realistic environments, road users must attend not only to the AV but also to surrounding traffic, raising the possibility that additional agents either contribute to visual noise (i.e., non-interacting, but attention-demanding stimuli) or create interference by directly intersecting the participant’s intended path. These interference situations constitute genuine interaction events in the sense of Markkula et al.~\cite{Markkula2020}, in which multiple agents compete for space and priority. Because both visual noise and interference can plausibly affect the detectability, interpretability, and perceived usefulness of an eHMI, we include them as systematic distraction conditions in our experimental design.

Additionally, the scenarios we chose focus specifically on yielding interactions (i.e., situations in which the AV decelerates and stops to allow the ego participant to proceed) because these moments represent the highest degree of ambiguity in AV/ road-user encounters. Prior research consistently shows that road users rely primarily on vehicle kinematics to infer intent and generally do so successfully in unambiguous approach or clearance phases~\cite{Dey2020_eHMI_gaze, dey2019gaze, dietrich2020automated, lee2021road}. However, when an AV begins to slow down, its behavioral cues become less diagnostic: deceleration can signal yielding, cautious approach, or uncertainty, making intent inherently ambiguous~\cite{declercq2019, habibovic2018communicating, Dey2020_eHMI_gaze}. This is precisely the interaction space where eHMIs have been found to be most influential, reducing uncertainty, improving predictability, and supporting trust when the vehicle’s intent cannot be inferred from kinematics alone~\cite{Dey2020_eHMI_gaze, LEE2022270, dietrich2020automated}. Given this prior evidence, we isolate yielding scenarios rather than the full driving cycle, as these represent the most theoretically and empirically relevant context for evaluating the added value of eHMIs.

\subsubsection{Pedestrian Scenarios}
\begin{figure}[ht]
    %\begin{subfigure}[b]{0.49\linewidth}
    %    \centering
    %    \includegraphics[width=\linewidth]{figures/PassantSzenarioNoise2.png}
    %    \caption{Schematic view.\newline}
    %    \label{fig:pedestrian-noise-schematic}
    %\end{subfigure}
    %\hfill
    %\begin{subfigure}[b]{0.49\linewidth}
        \centering
        \includegraphics[width=0.49\textwidth]{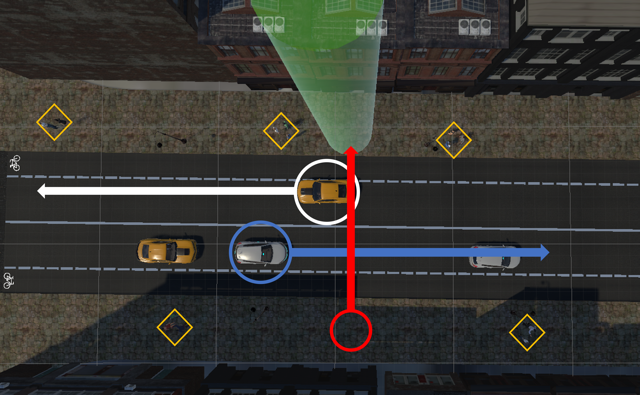}
        \caption{Noise scenario for the role of pedestrian. The yellow diamond-shaped rectangles indicate pedestrians standing at the street, adding visual noise, although not directly interfering with the participant's trajectory.}
    %    \label{fig:pedestrian-noise-unity}
    %\end{subfigure}
    %\caption{Top-down view on the pedestrian scenario with noise. The red represents the ego person, the blue the AV, and the black/white an MDV.}
    \label{fig:pedestrian-noise}
    \Description{Top-down view of the pedestrian scenario with noise. The red represents the ego person, the blue the AV, and the black/white an MDV.}
\end{figure}

\begin{figure}[ht]
        \centering
        \includegraphics[width=0.49\textwidth]{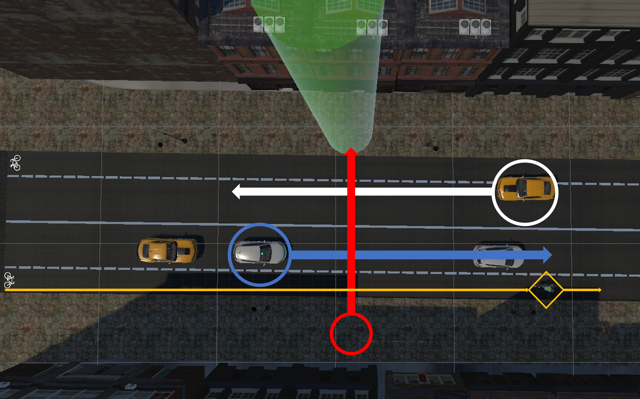}
        \caption{Interference scenario for the role of pedestrian. The yellow diamond-shaped rectangle indicates a cyclist crossing in front of the participant, directly interfering with their intended trajectory. }
    \label{fig:pedestrian-interference}
    \Description{Top-down view of the pedestrian scenario with interference.  The red represents the ego person, the blue is the AV, and the black/white is an MDV.}
\end{figure}

The pedestrian scenario starts at a sidewalk. The participant needs to cross a two-lane road. The road users in this scenario are MDVs, AVs, and the participant. On the lane closer to the participant, AVs and MDVs drive at a 50:50 rate. Only MDVs are on the far lane. \autoref{fig:pedestrian-noise} shows the scenario with noise, \autoref{fig:pedestrian-interference} with interference. 

After 18 seconds, the next AV on the near lane, from the perspective of the pedestrian (i.e., the participant), will stop and let the participant cross. This lets the participant experience multiple AVs and MDVs, representing an externally valid scenario.
%The timer was set to 18 seconds because the participant should see some MDVs and AVs that are not yielding, and that they can better understand the situation. Also, since the scenarios always have the same number and sequence of vehicles, at 18 seconds, two AVs are driving behind each other. The first one will still drive, while the second one yields for the participant. 
The MDVs on the far lane do not stop.
In the following figures, red represents the ego person, the blue the AV, and the black/white a MDV.

Visual noise is added to the scenario by adding standing pedestrians about 8 meters away from the participant. Therefore, these scenarios also represent scalability needs, adding realism to vehicle-pedestrian interaction scenarios by involving more than just two road users~\cite{10.1145/3334480.3382865}.

The interference scenario also extends the basic pedestrian scenario (see \autoref{fig:pedestrian-interference}). This scenario adds another road user (i.e., cyclist) with a conflict of trajectories with the participant, potentially causing an interference that could also happen in reality.

\subsubsection{Manual Driver Scenarios}
\label{subsection:driver-scenarios}

\begin{figure}[ht]
    %\begin{subfigure}[b]{0.40\linewidth}
    %    \centering
    %    \includegraphics[width=\linewidth]{figures/FahrerSzenarioNoise.png}
    %    \caption{Schematic view.}
    %    \label{fig:driver-noise-schematic}
    %\end{subfigure}
    %\hfill
    %\begin{subfigure}[b]{0.49\linewidth}
        \centering
        \includegraphics[width=0.49\textwidth]{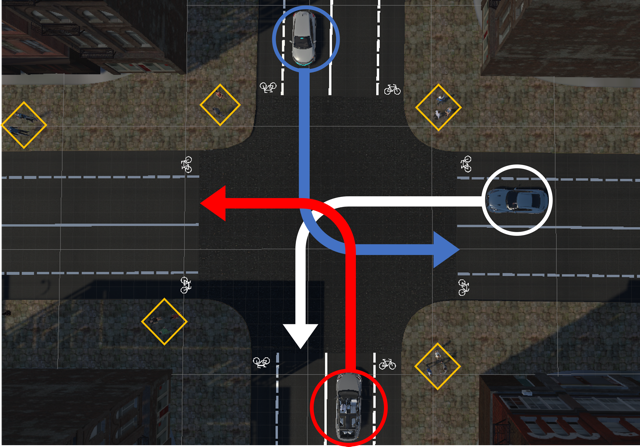}
        \caption{Noise scenario for the role of manual driver. The yellow diamond-shaped rectangles indicate pedestrians standing at the street, adding visual noise, although not directly interfering with the participant's trajectory.}
        %\label{fig:driver-noise-unity-schematic}
    %\end{subfigure}
    %\caption{Top-down view on the driver scenario with noise. The red represents the ego bicycle and the blue the AV.}
    \label{fig:driver-noise}
    \Description{Top-down view of the driver scenario with noise. The red represents the ego bicycle and the blue the AV.}
\end{figure}

\begin{figure}[ht]  
    %\begin{subfigure}[b]{0.40\linewidth}
    %    \centering
    %    \includegraphics[width=\linewidth]{figures/FahrerSzenarioInterference.png}
    %    \caption{Schematic view.}
    %    \label{fig:driver-interference-schematic}
    %\end{subfigure}
    %\hfill
    %\begin{subfigure}[b]{0.49\linewidth}
        \centering
        \includegraphics[width=0.49\textwidth]{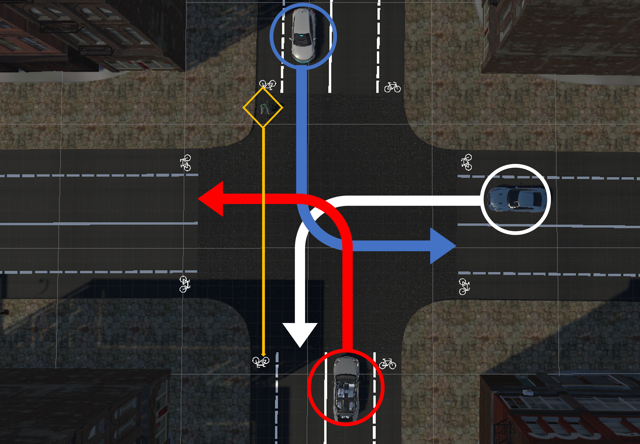}
        \caption{Interference scenario for the role of manual driver. The yellow diamond-shaped rectangles indicate cyclists crossing the street whose path directly interferes with the participant's trajectory.}
    %    \label{fig:driver-interference-unity-schematic}
    %\end{subfigure}
    %\caption{Top-down view on the driver scenario with interference.}
    \label{fig:driver-interference}
    \Description{Top-down view of the driver scenario with interference.}
\end{figure}

The manual driver scenario was adapted from \citet{10.1145/3546711}. It starts on a two-lane road (see \autoref{fig:driver-interference} and \autoref{fig:driver-noise}). The participant has to drive to the intersection and turn left. The finish zone is on the road after the participant turns left. The four-way intersection neither has signs nor traffic lights, so according to the StVO §8\footnote{\url{https://www.gesetze-im-internet.de/stvo_2013/__8.html}; accessed: 16.08.2025} Section 1 (right before left), the vehicle from the right (from the participant's point of view) has the right to go first. At the intersection, the participant (with the red rectangle, see \autoref{fig:driver-interference}) can spot two cars, one to the right and one in front. Both of them are signaling to turn left. The vehicle to the right is simulated to be manually driven, and the vehicle in front is an AV. Here, the German §9\footnote{\url{https://www.gesetze-im-internet.de/stvo_2013/__9.html}; accessed: 16.08.2025} section 3 StVO must also be applied. This means a vehicle must yield right of way to an oncoming vehicle if it wants to turn left. The result is that each vehicle has to wait for another vehicle. Thus, a deadlock occurs. This creates ambiguity and forces a careful interpretation and negotiation of the traffic situation.

Visual noise is again added to the scenario by adding standing pedestrians about 2-12 (depending on the participant's position) meters away from the participant.

The interference scenario adds cyclists crossing the participant's path. They drive on a bicycle path next to the lane with the AV. While the AV will still stop for the participant, the cyclist will not, as they have the right of way.

\subsubsection{Bicycle Scenarios}
\label{subsection:bicycle-scenarios}

\begin{figure}   
    %\begin{subfigure}[b]{0.40\linewidth}
    %    \centering
    %    \includegraphics[width=\linewidth]{figures/FahrradSzenarioNoise.png}
    %    \caption{Schematic view.\newline}
    %    \label{fig:bicycle-noise-schematic}
    %\end{subfigure}
    %\hfill
    %\begin{subfigure}[b]{0.45\linewidth}
        \centering
        \includegraphics[width=0.42\textwidth]{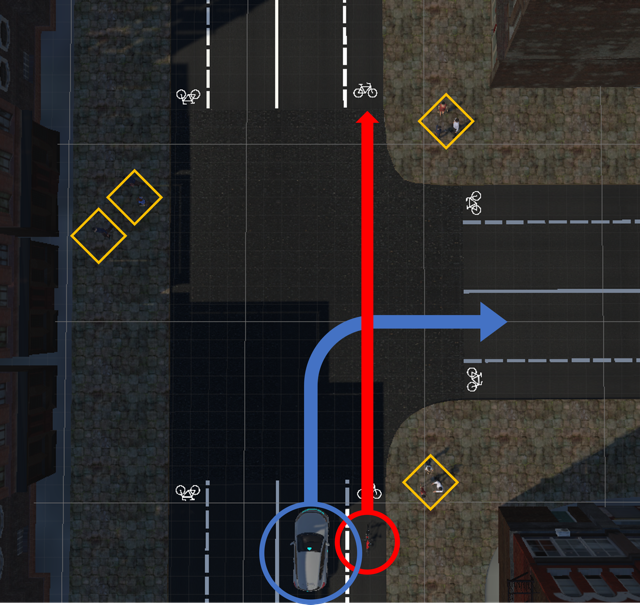}
        \caption{Noise scenario for the role of cyclist. The yellow diamond-shaped rectangles indicate pedestrians standing at the sidewalks of the intersection, adding visual noise, although not directly interfering with the participant's trajectory.}
    %    \label{fig:bicycle-noise-unity}
    %\end{subfigure}
    %\caption{Top-down view on the bicycle scenario with noise. The red represents the ego bicycle and the blue the AV.}
    \label{fig:bicycle-noise}
    \Description{Top-down view of the bicycle scenario with noise. The red represents the ego bicycle and the blue the AV.}
\end{figure}

The bicycle scenario starts with the participant on a bicycle path next to a two-lane road (see \autoref{fig:bicycle-noise}). The participant has to ride their bicycle straight. The finish zone is after a three-way intersection without traffic signs. In this scenario, only AVs drive on the right side of the two-lane road. The AVs turn right at the intersection, while the participant must go straight. The cyclist has the right of way as the AVs cross the participant's path.

Visual noise is added to the scenario via standing pedestrians about $1 m$ -- $20 m$ away from the participant (depending on their position and approach --- see \autoref{fig:bicycle-noise}).

\begin{figure}   
    %\begin{subfigure}[b]{0.40\linewidth}
    %    \centering
    %    \includegraphics[width=\linewidth]{figures/FahrradSzenarioInterference.png}
    %    \caption{Schematic view.\newline}
    %    \label{fig:bicycle-interference-schematic}
    %\end{subfigure}
    %\hfill
    %\begin{subfigure}[b]{0.49\linewidth}
        \centering
        \includegraphics[width=0.45\textwidth]{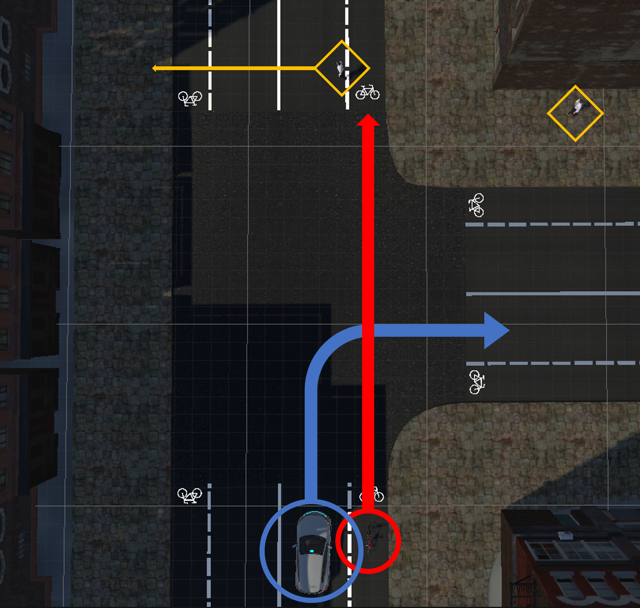}
        \caption{Interference scenario for the role of cyclist. The yellow diamond-shaped rectangles indicate interfering pedestrians crossing the street in a way that directly intersects the participant's trajectory on the bicycle.}
    %    \label{fig:bicycle-interference-unity-schematic}
    %\end{subfigure}
    %\caption{Top-down view on the driver scenario with interference. The red represents the ego vehicle, the blue the AV, and the white/black a MDV.}
    \label{fig:bicycle-interference}
    \Description{Top-down view of the driver scenario with interference. The red represents the ego vehicle, the blue the AV, and the white/black an MDV.}
\end{figure}

Interference is added via pedestrians walking over the street after the intersection (see \autoref{fig:bicycle-interference}).

\section{Experiment}
We designed and conducted a within-subject study with $N = 40$ participants to evaluate the effects of an eHMI on different road users (i.e., pedestrians, drivers, and cyclists) and the effect of different distraction levels (no distraction, visual noise, and interference) on the clarity or necessity of the eHMI. 
This research question (RQ) guided this study:

\begin{quote}
    \textit{RQ: What impact do the variables \textbf{eHMI}, \textbf{road user role} (i.e., manual driver, pedestrian, cyclist), and \textbf{distraction} have on road users in terms of (1) behavior, (2) mental demand, (3) trust and understanding, (4) perceived safety, (5) system usability, (6) usefulness, and satisfaction?}
\end{quote}

The within-subjects design allowed us to compare how the same individual interprets and evaluates an identical eHMI across three road-user roles without being constrained by the impractically large sample sizes to achieve comparable statistical sensitivity in a between-subjects design, especially given the three-way factorial structure. The within-subjects approach controls for inter-individual differences and allows for direct, participant-level comparisons of eHMI experiences across roles. To mitigate learning or carryover effects between role blocks, we employed full counterbalancing in the design. 

%1. Implicit vs Explicit eHMI: Are eHMIs necessary for these scenarios or these road users?

%2. For which road user(Pedestrian, Driver, Bicycle) is the communication with the eHMI of an AV most important?

%3. Does environmental distraction have an impact on communication with eHMIs?

Every participant experienced 18 conditions, resembling a 3 $\times$ 3 $\times$ 2 design. The independent variables were road user role (pedestrian, cyclist, driver), distraction (None, Noise, Interference), and eHMI (no and yes).

The experimental procedure followed the guidelines of our university's ethics committee and adhered to regulations regarding the handling of sensitive and private data, anonymization, compensation, and risk aversion. Compliant with our university’s local regulations, no additional formal ethics approval was required.

\subsection{Measurements}\label{sec:measurements}
\subsubsection{Objective Measurements} \label{objective-measurements}
\begin{figure*}[ht]
    \centering
    \begin{subfigure}[b]{0.33\textwidth}
        \centering
        \includegraphics[width=\linewidth]{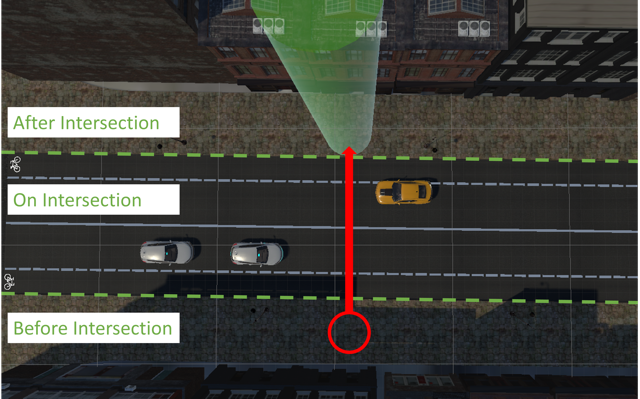}
        \caption{Pedestrian scenarios.}
        \label{fig:pedestrian-intersection-tracking}
    \end{subfigure}
    \hfill
    \begin{subfigure}[b]{0.33\textwidth}
        \centering
        \includegraphics[width=\linewidth]{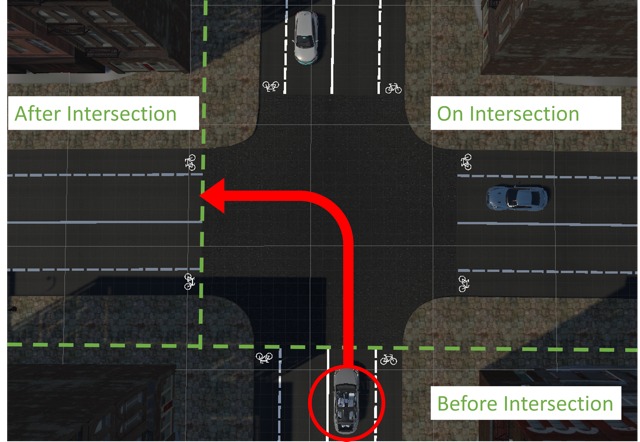}
        \caption{Manual drivers.}
        \label{fig:driver-intersection-tracking}
    \end{subfigure}
    \hfill
    \begin{subfigure}[b]{0.33\textwidth}
        \centering
        \includegraphics[width=\linewidth]{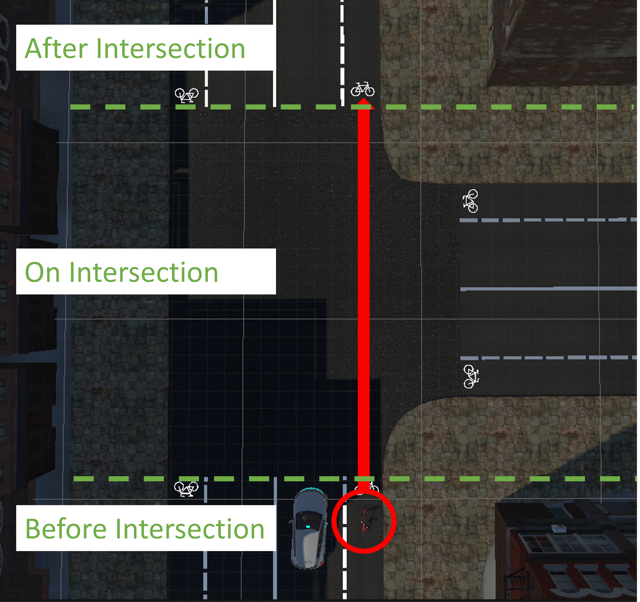}
        \caption{Cyclist scenarios.}
        \label{fig:bicycle-intersection-tracking}
    \end{subfigure}
    \caption{The three road users and their tracked areas.}
    \label{fig:intersection-tracking}
    \Description{The three road users and their tracked areas.}
\end{figure*}

We logged the participant's position and gaze at 50Hz. It also logged the time before the intersection, the time on the intersection, the time after the intersection, and the total duration (see \autoref{fig:intersection-tracking} for the areas). Also, since eye tracking was used, it tracked objects of interest if the participant was looking at them. These objects of interest are AVs, MDVs, noise objects (other pedestrians standing on the sidewalk), or interference objects (pedestrians crossing the street or cyclists riding on the street). If the participant was not looking at anything important to track, the value "null" was tracked.

\subsubsection{Subjective Measurements} \label{subjective-measurements}
We employed only the mental demand subscale of the raw NASA-TLX~\cite{hart1988development} on a 20-point scale (``How much mental and perceptual activity was required? Was the task easy or demanding, simple or complex?''; 1 = Very Low to 20 = Very High). 
Additionally, we used the subscales \textit{Predictability/Understandability} (hereafter \textit{Understandability}) and \textit{Trust} from the \textit{Trust in Automation} questionnaire by \citet{korber2018theoretical}. Understandability is measured using agreement on four statements (two direct: ``The system state was always clear to me.'', ``I was able to understand why things happened.''; and two inverse: ``The system reacts unpredictably.'', ``It's difficult to identify what the system will do next.'') using 5-point Likert scales (1 = Strongly disagree to 5 = Strongly agree). Trust is measured via agreement on equal 5-point Likert scales on two statements (``I trust the system.'' and ``I can rely on the system.'').
Also, participants rated their perceived safety using four 7-point semantic differentials from $-3$ (anxious/agitated/unsafe/timid) to +3 (relaxed/calm/safe/confident)~\cite{faas2020longitudinal}. Finally, participants stated their agreement to the self-developed statement ``The environment interfered in the communication with the automated vehicle'' (1 = Strongly disagree to 7=Strongly agree).

The system's usability was assessed with the System Usability Scale (SUS)~\cite{brooke1996sus} using a 5-point Likert scale (1=Strongly disagree to 5=Strongly agree).

Additionally, the van der Laan acceptance scale~\cite{VANDERLAAN19971} with the subscales "usefulness" and "satisfying" was employed.
The Acceptance scale is a simple way to assess the system's acceptance. It assesses System acceptance on a Usefulness scale and an affective satisfying scale using a nine-item Likert scale. % with the starting sentence "I find such a system ...". The Score is from +2 (useful/pleasant/bad /nice/effective/irritating/assisting /undesirable/raising alertness) to -2 (useless/unpleasant/ good/annoying/superfluous/ likeable/worthless/desirable/sleep-inducing). 
%Items 3, 6, and 8 are mirrored and should be scored -2 to +2. The Usefulness scale is the sum of items 1 + 3 + 5 + 7 + 9 divided by 5 (so it has a range from -2 to +2), and the Satisfying scale is the sum of items 2, 4, 6, and 8, divided by 4.

%own Question
As the last subjective measurement, a single item on a Likert scale from 1 ("Totally Disagree") to 7 ("Totally Agree") was used: "The environment interfered in the communication with the automated vehicle" was added.

\subsection{Procedure}
Each participant started by signing a declaration of consent. Following this, each participant received a short introduction to the study. Then, the VR headset was set up to sit comfortably on the participant's head. Also, we calibrated the eye tracking for every participant to ensure it worked properly. Subsequently, the participant read two texts depending on the scenario. The participants were only informed about the AVs and eHMI through the following texts. The intent of the eHMI was not explained to the participants to test for intuitiveness of the eHMI. The texts were translated from German to English and are shown in \autoref{app:intro_texts}. Participants were not told about what the eHMIs would convey, as we were interested in how intuitive they are.

Then, they experienced the scenario. Following the crossing task, participants completed a questionnaire containing subjective measures as described in Section~\ref{subjective-measurements}. The participants repeated reading the introductory texts, experiencing the scenarios, and answering the questionnaire until they finished all 18 scenarios. The study was fully counterbalanced. At the end of the study, the participants filled out a final questionnaire with demographic and open questions.

The study took approximately 75 minutes. Each cyclist scenario took \m{26.57} (\sd{6.55}), the driver scenario \m{22.05} (\sd{6.88}), and the pedestrian scenario \m{28.29} (\sd{3.42}) seconds.
Participants were compensated with 13€. %The study was conducted in English and German.

\section{Results}
\subsection{Data Analysis} \label{data-analysis}

While the scenarios per road user role differ, we still chose to compare the subjective dependent variables statistically. There are two arguments as to why we do so. First, we used the same eHMI and the same type of visual noise and interference, leading to some comparability. Secondly, the subjective dependent variables are less based on the scenario itself but on the interaction. For example, while the objective dependent variable \textit{total duration} is very dependent on the actual scenario and length of the journey, the subjective dependent variable \textit{trust} is mostly dependent on the interaction. This is also why we chose to only statistically evaluate the objective dependent variables, including the eHMI and distraction. 

Before every test, we checked all required assumptions. Since all data were nonparametric, we used the ARTool package by \citet{10.1145/1978942.1978963} and employed Holm correction for Dunn post-hoc tests. We used R in version 4.5.2 and RStudio in version 2025.09.2 and ensured that packages were up to date as of November 2025.
We report statistically significant main and interaction effects.

Control analyses, including trial number as a within-participant factor, did not reveal any significant main effects of trial for any dependent variable (see Appendix~\ref{app:order}), suggesting that learning or fatigue effects across the session were negligible in this dataset.

\subsection{Participants} \label{participants}
$N=40$ participants (21 female, 18 male, 1 non-binary) participated in the study. The participants were, on average, \m{25.075} (\sd{2.65}; range: 21 to 30) years old. Within the sample, 22 participants stated that their highest educational level was college, 16 high school, and 2 vocational training. Ten participants were employed, and 30 participants were college students.

On 5-point Likert scale (1 = Strongly Disagree - 5 = Strongly Agree), participants showed interest in AVs (\m{3.875}, \sd{1.22}), believed AVs to ease their lives (\m{3.775}, \sd{1.19}), and were unsure whether AVs become reality by 2033 (\m{3.35}, \sd{1.19}).

Participants were required to have normal or corrected-to-normal vision and to hold a valid driving license. No further exclusion criteria were defined; in particular, we did not screen based on prior VR experience, motion-sickness susceptibility, or cycling experience. All volunteers who consented and met these criteria were enrolled, and no additional candidates were excluded during screening.

\subsection{Subjective Measurements} \label{results-subjective-measurements}

\begin{table*}[htbp]
    \centering
    \small
    \caption{Summary of statistical main and interaction effects on dependent variables. (Dri: Driver, Cyc: Cyclist, Ped: Pedestrian).}
    \label{tab:results_summary}
    \renewcommand{\arraystretch}{1.2}
    \begin{tabular}{llcccl}
        \toprule
        \textbf{Measure} & \textbf{Factor} & \textbf{\textit{F}-value} & \textbf{\textit{p}-value} & \textbf{$\eta_{p}^{2}$} & \textbf{Post-hoc / Direction} \\
        \midrule
        
        % Mental Demand
        \multirow{3}{*}{Mental Demand} 
        & Role & $F(2,78) = 3.57$ & .033 & .08 & Cyc $>$ Ped \\
        & Distraction & $F(2,78) = 4.28$ & .017 & .10 & n.s. \\
        & eHMI & $F(1,39) = 35.23$ & $<.001$ & .47 & No eHMI $>$ eHMI \\
        \midrule
        
        % Perceived Safety
        \multirow{2}{*}{Perceived Safety} 
        & Role & $F(2,78) = 6.63$ & .002 & .15 & Dri $>$ Cyc; Ped $>$ Cyc \\
        & eHMI & $F(1,39) = 47.88$ & $<.001$ & .55 & eHMI $>$ No eHMI \\
        \midrule
        
        % Usability
        \multirow{4}{*}{Usability (SUS)} 
        & Role & $F(2,78) = 5.84$ & .004 & .13 & Ped (Highest), Cyc (Lowest) \\
        & eHMI & $F(1,39) = 37.19$ & $<.001$ & .49 & eHMI $>$ No eHMI \\
        & Role $\times$ Dist. & $F(4,156) = 2.78$ & .029 & .07 & -- \\
        & Role $\times$ Dist. $\times$ eHMI & $F(4,156) = 2.60$ & .038 & .06 & -- \\
        \midrule
        
        % Usefulness
        \multirow{2}{*}{Usefulness} 
        & Role & $F(2,78) = 6.93$ & .002 & .15 & Dri $>$ Cyc; Ped $>$ Cyc \\
        & eHMI & $F(1,39) = 72.13$ & $<.001$ & .65 & eHMI $>$ No eHMI \\
        \midrule
        
        % Satisfaction
        \multirow{2}{*}{Satisfaction} 
        & Role & $F(2,78) = 8.39$ & $<.001$ & .18 & Ped $>$ Cyc \\
        & eHMI & $F(1,39) = 55.69$ & $<.001$ & .59 & eHMI $>$ No eHMI \\
        \midrule
        
        % Trust
        \multirow{3}{*}{Trust} 
        & Role & $F(2,78) = 6.57$ & .002 & .14 & Ped $>$ Cyc; Ped $>$ Dri \\
        & Distraction & $F(2,78) = 6.94$ & .002 & .15 & n.s. \\
        & eHMI & $F(1,39) = 40.96$ & $<.001$ & .51 & eHMI $>$ No eHMI \\
        \midrule
        
        % Understandability
        \multirow{3}{*}{Understandability} 
        & Role & $F(2,78) = 3.66$ & .030 & .09 & Ped $>$ Cyc \\
        & Distraction & $F(2,78) = 4.49$ & .014 & .10 & n.s. \\
        & eHMI & $F(1,39) = 48.09$ & $<.001$ & .55 & eHMI $>$ No eHMI \\
        \midrule
        
        % Interference
        Interference & Distraction & $F(2,78) = 3.98$ & .023 & .09 & n.s. \\
        
        \bottomrule
    \end{tabular}
\end{table*}

\subsubsection{Mental Demand, Perceived Safety}\label{sec:mw_ps}
%TLX1
The ART found a significant main effect of the road user role (\F{2}{78}{3.57}, \p{0.033}, $\eta_{p}^{2}$ = 0.08, 95\% CI: [0.00, 1.00]), of Distraction (\F{2}{78}{4.28}, \p{0.017}, $\eta_{p}^{2}$ = 0.10, 95\% CI: [0.01, 1.00]), and of eHMI on mental demand (\F{1}{39}{35.23}, \pminor{0.001}, $\eta_{p}^{2}$ = 0.47, 95\% CI: [0.28, 1.00]).
A post-hoc test found that the cyclists were significantly higher (\m{9.51}, \sd{4.75}) in terms of mental demand compared to the pedestrians (\m{8.47}, \sd{4.67}; \padj{0.037}, \rankbiserial{0.14}). 
A post-hoc test found no significant differences in distraction.
Additionally, the mental demand in scenarios without eHMI (\m{10.24}, \sd{4.60}) was significantly higher compared to scenarios with the intention eHMI (\m{7.81}, \sd{4.55}; \rankbiserial{0.30}).

%Perceived Safety
The ART found a significant main effect of the road user role (\F{2}{78}{6.63}, \p{0.002}, $\eta_{p}^{2}$ = 0.15, 95\% CI: [0.04, 1.00]) and of eHMI on perceived safety (\F{1}{39}{47.88}, \pminor{0.001}, $\eta_{p}^{2}$ = 0.55, 95\% CI: [0.37, 1.00]). 
A post-hoc test found that the drivers were significantly higher (\m{1.28}, \sd{1.35}) in terms of perceived safety compared to the cyclists (\m{0.87}, \sd{1.43}; \padj{0.003}, \rankbiserial{0.17}). The test also found that the pedestrians were significantly higher (\m{1.23}, \sd{1.41}) in terms of perceived safety compared to the cyclists (\m{0.87}, \sd{1.43}; \padj{0.008}, \rankbiserial{0.15}). The participant felt significantly safer in scenarios with an eHMI (\m{1.62}, \sd{1.17}) compared to those without the eHMI (\m{0.64}, \sd{1.46}, \rankbiserial{-0.40}).

\subsubsection{Usability}
\begin{figure}[ht!]
    \centering
    \includegraphics[width=0.5\textwidth]{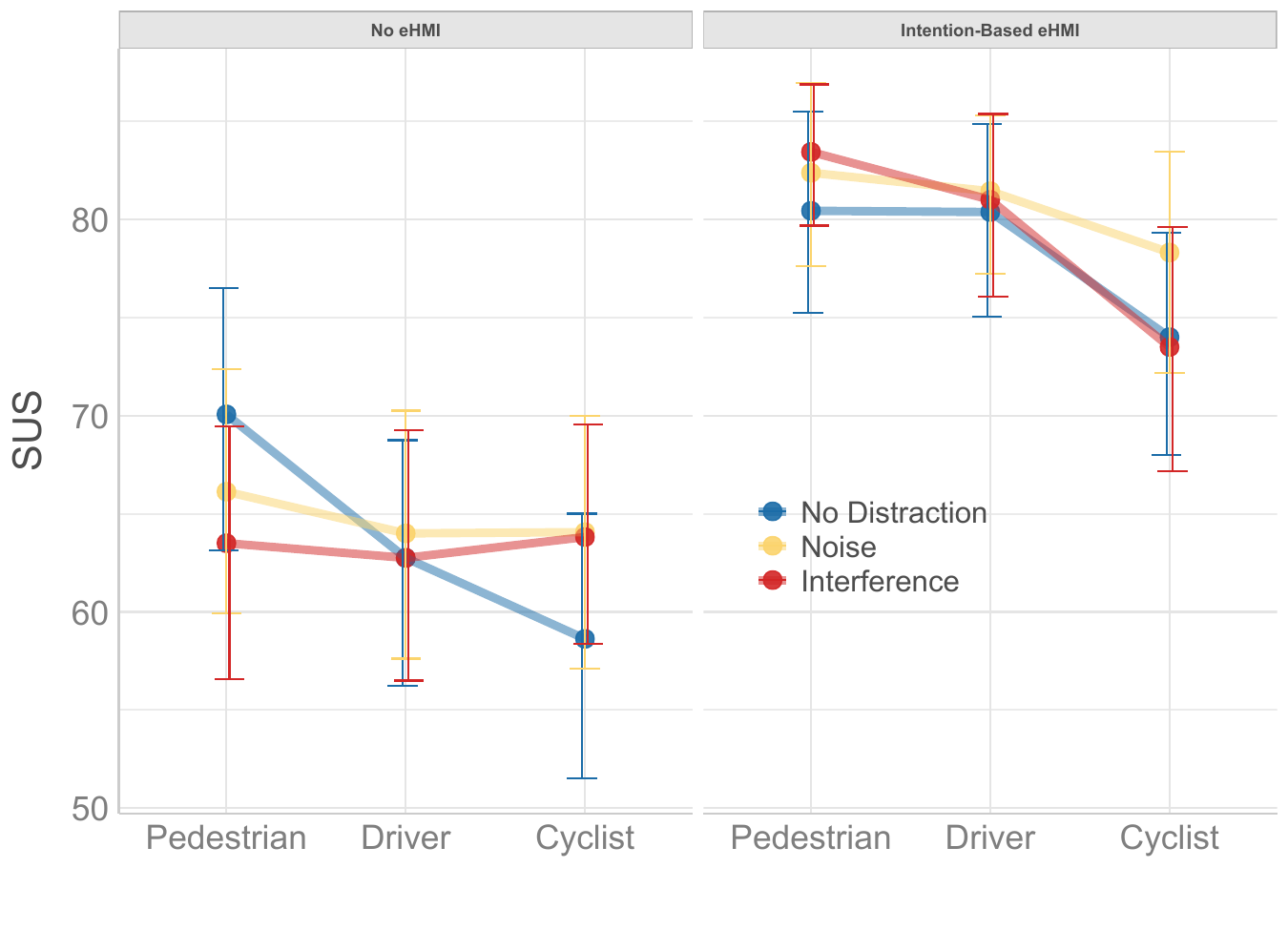}
    \caption{The three-way interaction effect of the road user role $\times$ Distraction $\times$ eHMI on the SUS score. Error bars represent 95\% confidence intervals.}
    \label{fig:sus-3way}
    \Description{The three-way interaction effect of the road user role times Distraction times eHMI on the SUS score.}
\end{figure}

%SUS
The ART found a significant main effect of the road user role (\F{2}{78}{5.84}, \p{0.004}, $\eta_{p}^{2}$ = 0.13, 95\% CI: [0.03, 1.00]) and of eHMI on system usability (\F{1}{39}{37.19}, \pminor{0.001}, $\eta_{p}^{2}$ = 0.49, 95\% CI: [0.30, 1.00]). The ART found a significant interaction effect of the road user role $\times$ Distraction on system usability (\F{4}{156}{2.78}, \p{0.029}, $\eta_{p}^{2}$ = 0.07, 95\% CI: [0.00, 1.00]) and also a significant interaction effect of the road user role $\times$ Distraction $\times$ eHMI on system usability (\F{4}{156}{2.60}, \p{0.038}, $\eta_{p}^{2}$ = 0.06, 95\% CI: [0.00, 1.00]; see \autoref{fig:sus-3way}).

%\autoref{fig:sus-3way} shows the three-way interaction effect of the road user role $\times$ Distraction $\times$ eHMI on system usability. 
On the System Usability Score (SUS), anything above 68 is considered above average~\cite{SUSMeasuring}. So, while most of the scenarios without eHMI scored an SUS score below 68, every Scenario with an eHMI scored an SUS score over 68. Also, the SUS for cyclists was rated the worst (especially when there was no distraction), while the SUS was rated the highest for pedestrians. In general, the SUS score followed the same pattern for no distraction, noise, and interference.

\subsubsection{Usefulness and Satisfaction}
%AOAUsefuleness
The ART found a significant main effect of the road user role (\F{2}{78}{6.93}, \p{0.002}, $\eta_{p}^{2}$ = 0.15, 95\% CI: [0.04, 1.00]) and of eHMI on usefulness (\F{1}{39}{72.13}, \pminor{0.001}, $\eta_{p}^{2}$ = 0.65, 95\% CI: [0.49, 1.00]). 

A post-hoc test found that usefulness for drivers was significantly higher (\m{0.53}, \sd{0.82}) than for cyclists (\m{0.37}, \sd{0.82}; \padj{0.042}, \rankbiserial{0.13}) as well as that usefulness for pedestrian was significantly higher (\m{0.63}, \sd{0.82}) than for cyclists (\m{0.37}, \sd{0.82}; \padj{0.001}, \rankbiserial{0.18}). 

Participants assessed scenarios with eHMIs (\m{0.91}, \sd{0.72}) significantly more useful than scenarios without eHMI (\m{0.11}, \sd{0.73}, \rankbiserial{-0.58}).

%AOASatisfying
%The ART found a significant main effect of Scenario on satisfaction (\F{2}{78}{8.39}, \pminor{0.001}). The ART found a significant main effect of eHMI on satisfaction (\F{1}{39}{55.69}, \pminor{0.001}).
The ART found a significant main effect of the road user role (\F{2}{78}{8.39}, \pminor{0.001}, $\eta_{p}^{2}$ = 0.18, 95\% CI: [0.06, 1.00]), and of eHMI on satisfaction (\F{1}{39}{55.69}, \pminor{0.001}, $\eta_{p}^{2}$ = 0.59, 95\% CI: [0.42, 1.00]).
%A post-hoc test found that the cyclists were significantly higher (\m{-0.26}, \sd{1.01}) in terms of AOASatisfying compared to Pedestrian (\m{-0.58}, \sd{0.96}; \padj{0.002}). 
A post-hoc test found that the pedestrians were significantly higher (\m{0.58}, \sd{0.96}) in terms of satisfaction compared to the cyclists (\m{0.26}, \sd{1.01}; \padj{0.002}, \rankbiserial{0.18}).
%No EHMI: \m{0.04}, \sd{0.93}
%Intention: \m{-0.88}, \sd{0.82}
Also, participants assessed satisfaction as significantly higher in scenarios with eHMI (\m{0.88}, \sd{0.82}) compared to those without eHMI (\m{-0.04}, \sd{0.93}, \rankbiserial{-0.55}).

%Trust
\subsubsection{Trust and Understandability}
The ART found a significant main effect of the road user role (\F{2}{78}{6.57}, \p{0.002}, $\eta_{p}^{2}$ = 0.14, 95\% CI: [0.04, 1.00]), of Distraction (\F{2}{78}{6.94}, \p{0.002}, $\eta_{p}^{2}$ = 0.15, 95\% CI: [0.04, 1.00]), and of eHMI on trust (\F{1}{39}{40.96}, \pminor{0.001}, $\eta_{p}^{2}$ = 0.51, 95\% CI: [0.33, 1.00]).

A post-hoc test found that Trust for the \Scenario Pedestrian was significantly higher (\m{3.53}, \sd{1.07}) than for Cyclist (\m{3.21}, \sd{1.11}; \padj{0.003}, \rankbiserial{0.17}) and Driver (\m{3.29}, \sd{1.08}; \padj{0.032}, \rankbiserial{0.13}).
A post-hoc test found no significant differences in trust regarding Distraction. 
The participants' trust score was higher for scenarios with eHMI (\m{3.72}, \sd{1.05}) compared to those without eHMIs (\m{2.97}, \sd{1.02}, \rankbiserial{-0.41}).

%TrustUnderstanding
The ART found a significant main effect of the road user role (\F{2}{78}{3.66}, \p{0.030}, $\eta_{p}^{2}$ = 0.09, 95\% CI: [0.00, 1.00]), of Distraction (\F{2}{78}{4.49}, \p{0.014}, $\eta_{p}^{2}$ = 0.10, 95\% CI: [0.01, 1.00]), and of eHMI on Understandability (\F{1}{39}{48.09}, \pminor{0.001}, $\eta_{p}^{2}$ = 0.55, 95\% CI: [0.37, 1.00]). 
A post-hoc test found that the pedestrians were significantly higher (\m{3.61}, \sd{1.07}) in terms of Understandability compared to the cyclists (\m{3.37}, \sd{1.08}; \padj{0.039}, \rankbiserial{0.13}). 
A post-hoc test found no significant differences in Understandability regarding Distraction. 
Also, the scenarios with eHMI scored higher in Understandability (\m{3.93}, \sd{0.93}) than the scenarios without eHMI (\m{3.11}, \sd{1.03}, \rankbiserial{-0.45}).

\subsubsection{Interference of Environment} \label{Interference of environment}
The ART found a significant main effect of Distraction on the interference of the environment (\F{2}{78}{3.98}, \p{0.023}, $\eta_{p}^{2}$ = 0.09, 95\% CI: [0.01, 1.00]).
However, a post-hoc test found no significant differences.

\subsection{Objective Measurements}\label{results-objective-measurements}
We statistically compared the time before entering the street (for pedestrians) or entering the intersection (for drivers and cyclists) and the total duration. For eye-tracking, we only describe the percentages per road user role. 

\subsubsection{Road User Role: Pedestrian}

\paragraph{Time Before Entering Street and Total Duration}
The ART found a significant main effect of distraction on time before entering the street (\F{2}{78}{12.20}, \pminor{0.001}, $\eta_{p}^{2}$ = 0.24, 95\% CI: [0.10, 1.00]). 
A post-hoc test found that Interference was significantly higher (\m{18.08}, \sd{6.95}) in terms of time before entering street compared to  Noise (\m{13.40}, \sd{9.02}; \padj{0.022}, \rankbiserial{0.25}) and None (\m{14.37}, \sd{9.02}; \padj{0.024}, \rankbiserial{0.23}). The ART found no significant effects on total duration.

\paragraph{Eye-Tracking}

\begin{figure}
    \centering
    \includegraphics[width=0.5\textwidth]{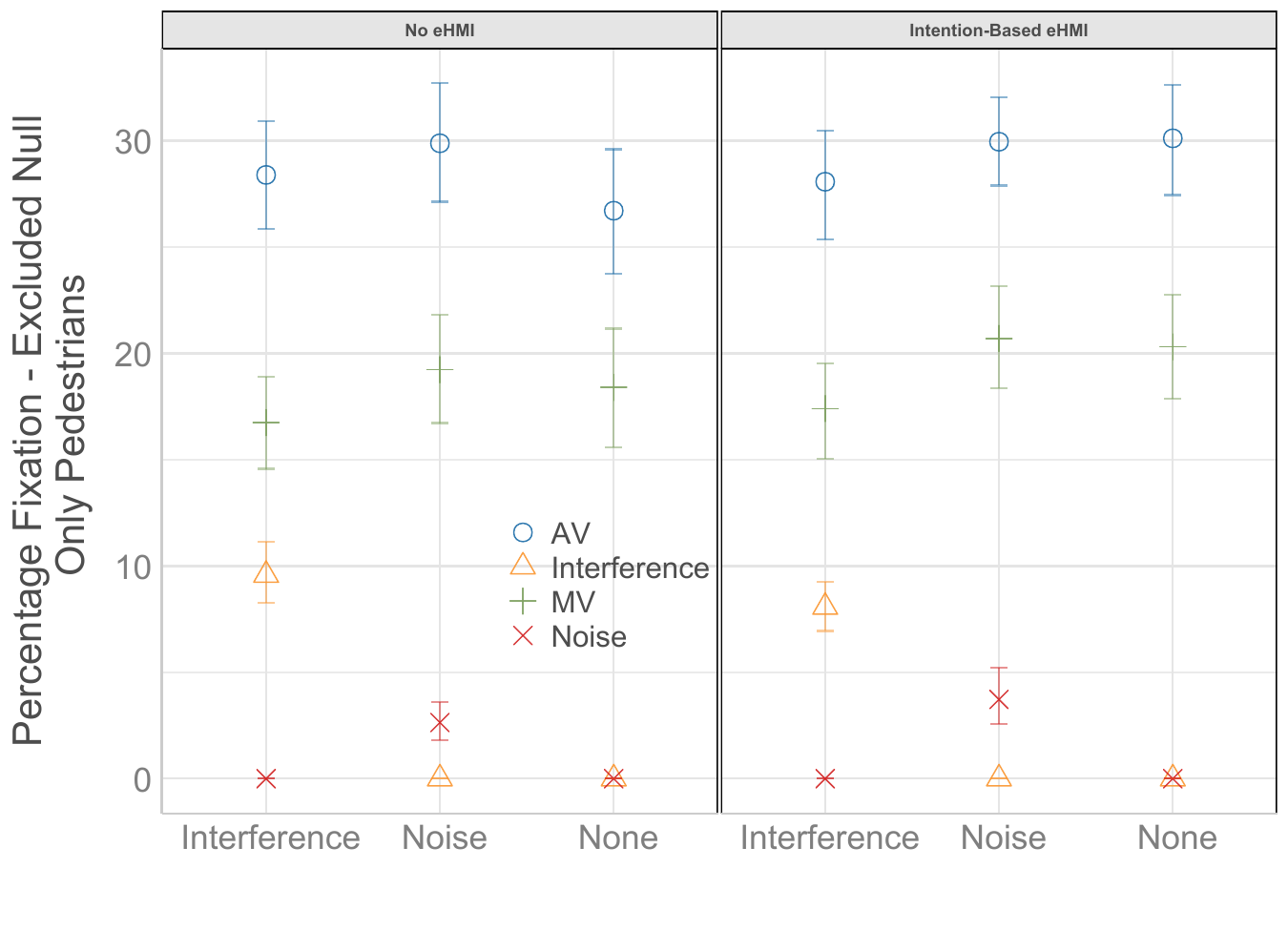}
    \caption{The percentage of AOIs for all pedestrians.}
    \label{fig:eye-tracking-pedestrian}
    \Description{The percentage of AOIs for all pedestrians.}
\end{figure}

\autoref{fig:eye-tracking-pedestrian} shows the eye-tracking data for all pedestrians. The distribution of percentages is equal for those with and without an eHMI. AVs were looked at the most, followed by manual vehicles.

\subsubsection{Road User Role: Manual Driver}

\paragraph{Time Before Entering Intersection and Total Duration}
%No eHMI: \m{15.05}, \sd{5.74}
%Intention-Based eHMI: \m{13.79}, \sd{5.51}
The ART found a significant main effect of eHMI on time before entering the intersection (\F{1}{39}{8.20}, \p{0.007}, $\eta_{p}^{2}$ = 0.17, 95\% CI: [0.03, 1.00]). Without an eHMI (\m{15.05}, \sd{5.74}), participants needed significantly longer than with (\m{13.79}, \sd{5.51}, \rankbiserial{0.16}).
%No eHMI: \m{22.74}, \sd{6.90}
%Intention-Based eHMI: \m{21.36}, \sd{6.82}
The ART found a significant main effect of eHMI on total duration (\F{1}{39}{7.39}, \p{0.010}, $\eta_{p}^{2}$ = 0.16, 95\% CI: [0.02, 1.00]). 
Without an eHMI (\m{22.74}, \sd{6.90}), participants needed significantly longer than with (\m{21.36}, \sd{6.82}, \rankbiserial{0.14}).

\paragraph{Eye-Tracking}

\begin{figure}
    \centering
    \includegraphics[width=1.00\linewidth]{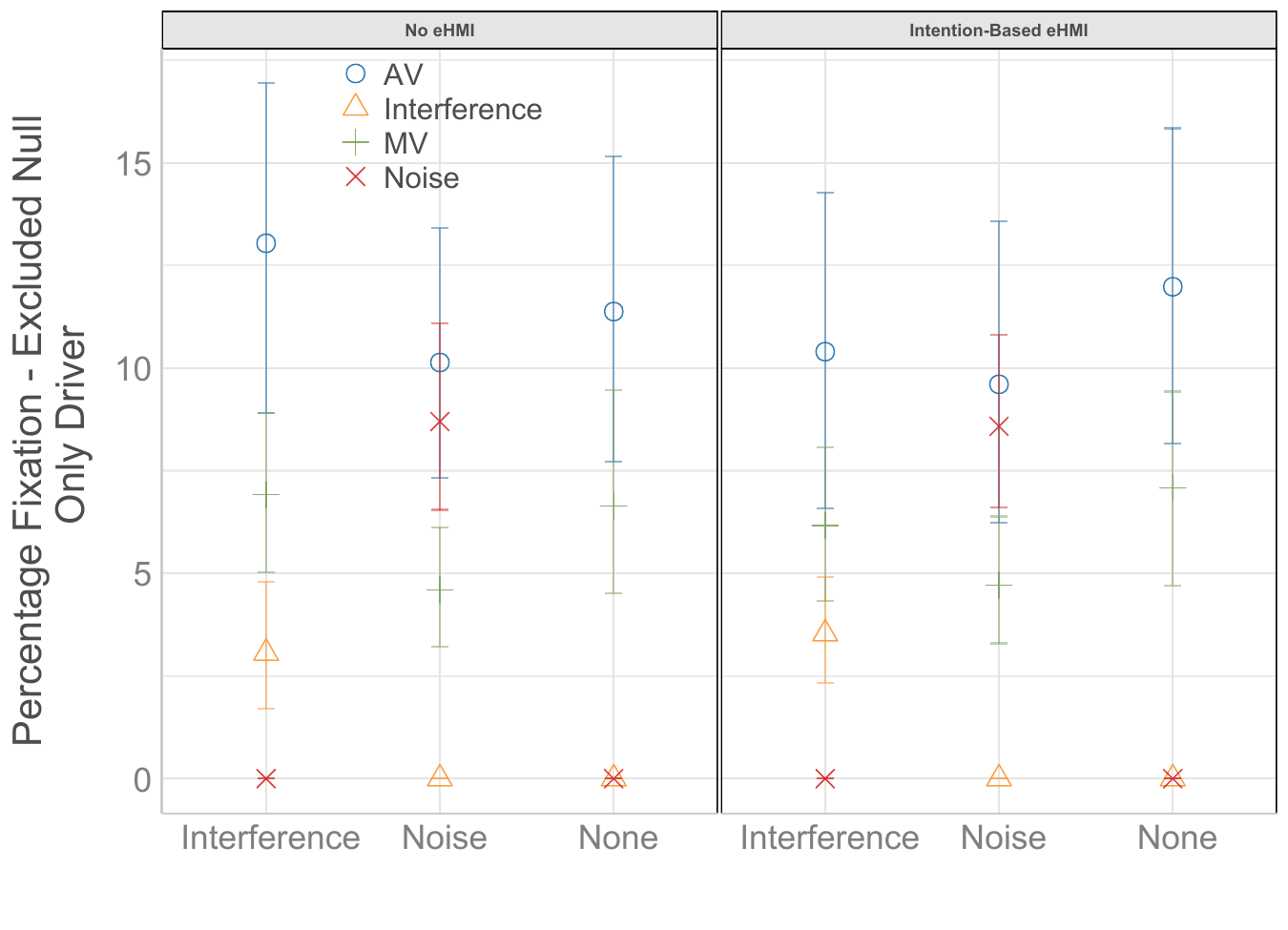}
    \caption{The percentage of AOIs for all drivers.}
    \label{fig:eye-tracking-driver}
    \Description{The percentage of AOIs for all drivers.}
\end{figure}

\autoref{fig:eye-tracking-driver} shows the eye-tracking data for all drivers. The distribution of percentages is equal for those with and without an eHMI. AVs were looked at the most, followed by manual vehicles. If there was noise, this was, however, more focused than the manual vehicles.

\subsubsection{Road User Role: Cyclist}

\paragraph{Time Before Entering Street and Total Duration}

%No eHMI: \m{18.84}, \sd{5.06}
%Intention-Based eHMI: \m{20.24}, \sd{6.51}
The ART found a significant main effect of eHMI on time before entering the intersection (\F{1}{39}{4.97}, \p{0.032}, $\eta_{p}^{2}$ = 0.11, 95\% CI: [0.01, 1.00]). Here, without an eHMI (\m{18.84}, \sd{5.06}), participants needed \textbf{less} time than with (\m{20.24}, \sd{6.51}, \rankbiserial{-0.11}). The ART found no significant effects on total duration.

\paragraph{Eye-Tracking}

\begin{figure}
    \centering
    \includegraphics[width=1.00\linewidth]{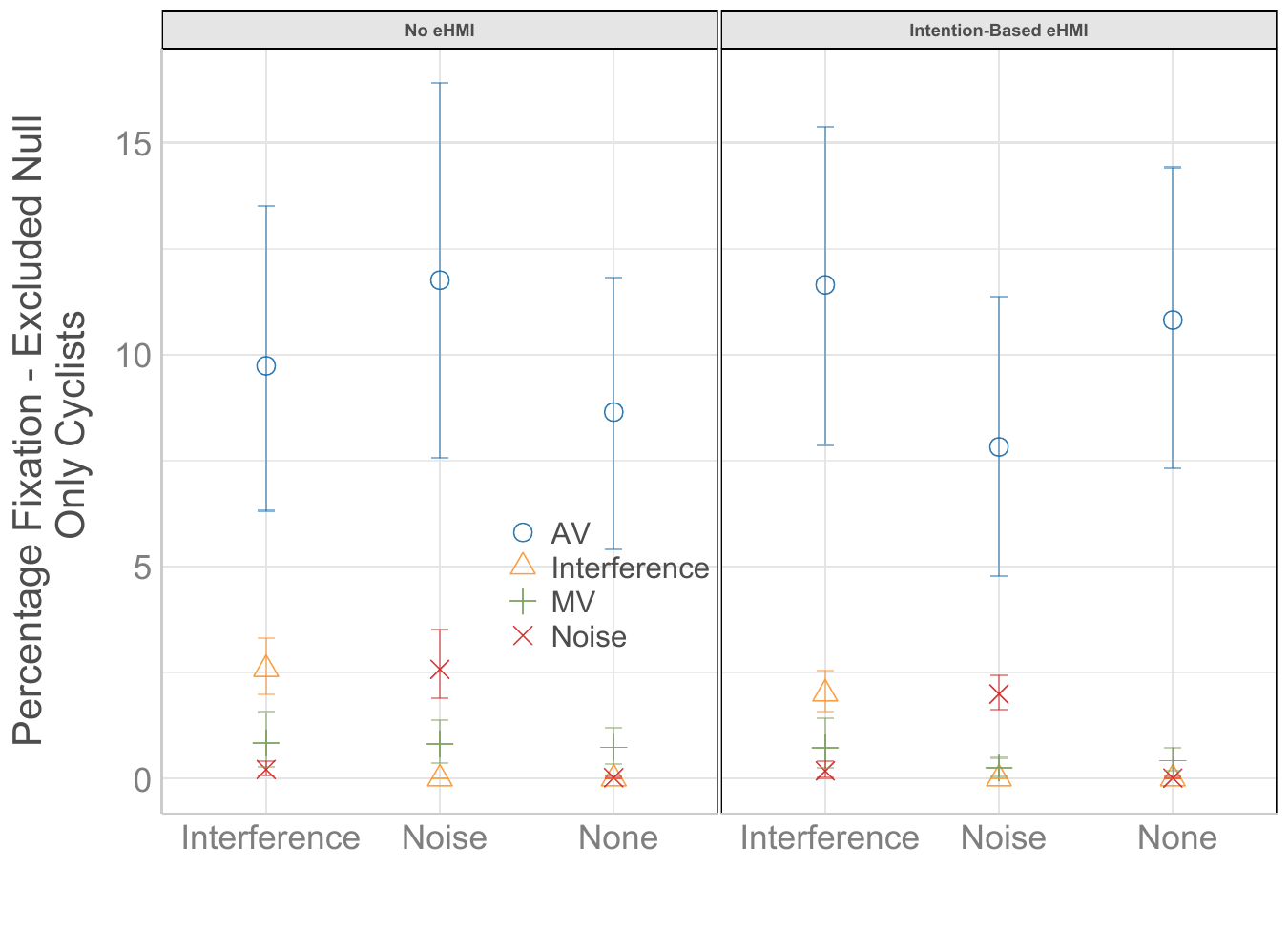}
    \caption{The percentage of AOIs for all cyclist scenarios.}
    \label{fig:eye-tracking-cyclist}
    \Description{The percentage of AOIs for all cyclist scenarios.}
\end{figure}

\autoref{fig:eye-tracking-cyclist} shows the eye-tracking data for all cyclist scenarios. The distribution of the percentages is equal for having an eHMI or not - only for looking at AVs is it different. AVs were looked at most; however, when having the eHMI, the AVs were looked at more for interference and scenarios without distraction.

\subsection{Time to First Fixation (TTFF)}

\begin{figure*}[ht!]
    \centering
    \includegraphics[width=\textwidth]{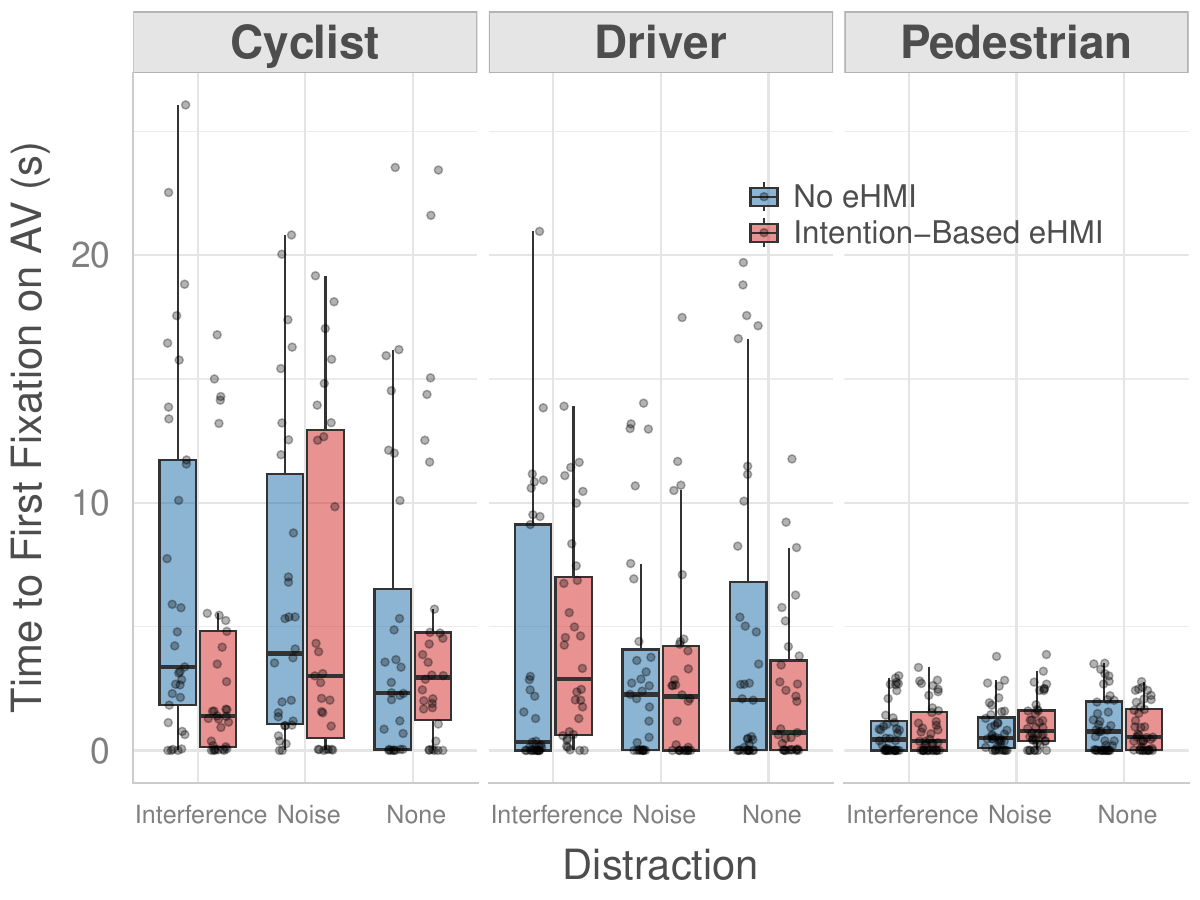}
    \caption{Time to First Fixation for all conditions.}
    \label{fig:eye-tracking-ttff}
    \Description{Time to First Fixation for all conditions.}
\end{figure*}

We fitted a linear mixed model (estimated using REML and nloptwrap optimizer) to predict ttff with distraction, ehmi, and scenario (formula: ttff $\sim$ distraction * ehmi * scenario). The model included participant as random effect (formula: $\sim$1 | participant). The model's total explanatory power is substantial (conditional $R^2$ = 0.30) and the part related to the fixed effects alone (marginal $R^2$) is of 0.19. The model's intercept, corresponding to distraction =  Interference, ehmi = No eHMI and scenario = Cyclist, is at 7.04 (95\% CI [5.53, 8.55], t(593) = 9.16, p < .001). Within this model:

\begin{itemize}
\item The effect of ehmi [Intention-Based eHMI] is statistically significant and negative (beta = -3.39, 95\% CI [-5.39, -1.38], t(593) = -3.32, p < .001; Std. beta = -0.70, 95\% CI [-1.11, -0.28])
\item The effect of scenario [Driver] is statistically significant and negative (beta = -3.40, 95\% CI [-5.40, -1.39], t(593) = -3.33, p < .001; Std. beta = -0.70, 95\% CI [-1.11, -0.29])
\item The effect of scenario [Pedestrian] is statistically significant and negative (beta = -6.21, 95\% CI [-8.12, -4.30], t(593) = -6.37, p < .001; Std. beta = -1.28, 95\% CI [-1.67, -0.88])
\item The effect of distraction [ Noise] × ehmi [Intention-Based eHMI] is statistically significant and positive (beta = 3.34, 95\% CI [0.38, 6.29], t(593) = 2.22, p = 0.027; Std. beta = 0.69, 95\% CI [0.08, 1.29])
\item The effect of distraction [ None] × ehmi [Intention-Based eHMI] is statistically significant and positive (beta = 3.63, 95\% CI [0.70, 6.56], t(593) = 2.43, p = 0.015; Std. beta = 0.75, 95\% CI [0.14, 1.35])
\item The effect of distraction [ None] × scenario [Driver] is statistically significant and positive (beta = 3.09, 95\% CI [0.22, 5.97], t(593) = 2.11, p = 0.035; Std. beta = 0.64, 95\% CI [0.04, 1.23])
\item The effect of ehmi [Intention-Based eHMI] × scenario [Driver] is statistically significant and positive (beta = 4.29, 95\% CI [1.44, 7.14], t(593) = 2.96, p = 0.003; Std. beta = 0.88, 95\% CI [0.30, 1.47])
\item The effect of ehmi [Intention-Based eHMI] × scenario [Pedestrian] is statistically significant and positive (beta = 3.43, 95\% CI [0.73, 6.14], t(593) = 2.49, p = 0.013; Std. beta = 0.71, 95\% CI [0.15, 1.26])
\item The effect of (distraction [ Noise] × ehmi [Intention-Based eHMI]) × scenario [Driver] is statistically significant and negative (beta = -4.66, 95\% CI [-8.80, -0.52], t(593) = -2.21, p = 0.027; Std. beta = -0.96, 95\% CI [-1.81, -0.11])
\item The effect of (distraction [ None] × ehmi [Intention-Based eHMI]) × scenario [Driver] is statistically significant and negative (beta = -6.79, 95\% CI [-10.87, -2.70], t(593) = -3.26, p = 0.001; Std. beta = -1.40, 95\% CI [-2.24, -0.56])
\end{itemize}

Standardized parameters were obtained by fitting the model on a standardized version of the dataset. 95\% Confidence Intervals (CIs) and p-values were computed using a Wald t-distribution approximation.

\subsection{Detailed Gaze Metrics and Relationship to Subjective Measures}

To provide deeper insight into the allocation of visual attention beyond AOI percentages, we calculated temporal gaze metrics for the AV AOI. These included Time to First Fixation (TTFF), serving as a measure of attentional capture (how quickly the AV was noticed), and Total Fixation Duration, serving as a measure of information processing depth.

We performed Spearman’s rank correlations to investigate the relationship between these objective gaze metrics and the participants' subjective ratings (Trust, Understanding, Perceived Safety, and Mental Demand).

\paragraph{Visual Attention and Understanding}
Contrary to the hypothesis that increased visual attention facilitates higher understanding, we found no significant correlation between the Total Fixation Duration on the AV and the self-reported Understanding (subscale of TiA; $\rho$ = 0.03, S = 3.72e+07, p =0.462).
Similarly, the Average Fixation Duration (the average length of a single glance) did not correlate with overall Trust ($\rho$ = 3.81e-03, S = 3.82e+07, p = 0.925). This suggests that participants did not necessarily require prolonged gaze durations to achieve high subjective understanding or trust levels; rather, efficient visual checks may have been sufficient.

\paragraph{Visual Attention and Mental Demand}
A significant negative correlation was found between Total Fixation Duration on the AV and Mental Demand (NASA-TLX; $\rho$ = -0.11, S = 4.24e+07, p = 0.009). This indicates that participants who allocated more total time to observing the AV generally reported lower mental demand. This finding suggests that the AV (and its eHMI) served as a supportive information source—looking at it more frequently or for longer periods was associated with a reduction in perceived cognitive load.

\paragraph{Visual Attention and Safety}
We examined whether noticing the AV earlier (lower TTFF) was associated with higher perceived safety. The Spearman correlation revealed no significant relationship between TTFF and Perceived Safety ($\rho$ = -0.04, S = 3.98e+07, p = 0.381).

%%%%%%%%%%%%%%%%%%%%%%%%%%%%%%%%%%%%%%%%%%%%%%%%%%%%%%%%%%%%%%%%%%%

\subsection{Subjective Insights and Improvement Proposals}
We did not perform a formal analysis of the participants' qualitative feedback. Instead, we present anecdotal quotes below, summarizing the key points shared by the participants.
Quotes were given in German and English, but were translated into English.

%\subsubsection{Key Insights}
\paragraph{Visibility and Clarity}
Participants generally found the communication via the SPLB eHMI and the indication of it being an AV helpful, yet emphasized the need for clear differentiation from standard vehicle signals.
\begin{quote}
``The blinking was nice and pleasant and I liked that it was a cool blue color that didn't bother my eyes and was very clear but would maybe need an explanation that blinking means the AV waits.''
\end{quote}

\paragraph{Timing and Predictability}
Effective communication was often hampered by the timing of the signals. Participants noted that AVs sometimes communicated their intentions too late.

\begin{quote}
``Maybe to solve the problem I described, where the eHMIs would only be used at the last second, I would propose to use some kind of pre-signal.'' 
\end{quote}

It was also interesting to see that one participant perceived the general intention-display later with an eHMI than without (i.e., only communicating via deceleration):
\begin{quote}
``They reacted differently than expected, they always showed their intentions very late when they were equipped with the HMI.'' 
\end{quote}

\paragraph{Color and Signal Interpretation}
The choice of signal color and type was crucial. While cyan was generally pleasant, some found it unclear and proposed to use more generally understandable colors (e.g., as learned from traffic lights).
\begin{quote}
``Use a different color that is better to see on a white car (e.g., amber).'' 
\end{quote}
\begin{quote}
%"Das System sollte in einer auffälligeren Farbe als einem hellblau sein. Rot wär nicht schlecht." / `
`The system should be in a more noticeable color than light blue. Red wouldn't be bad.''
\end{quote}

However, proposed colors such as amber carry meaning in the context of traffic and, therefore, should be avoided~\cite{dey2020color}.

\paragraph{Differing Experiences Across User Types}
\begin{itemize}
    \item \textbf{Pedestrians:} ``I felt a lot more safe while the vehicles indicated their behavior with me, especially as pedestrian.''
    \item \textbf{Cyclists:} ``It was mostly helpful and always a nice thing to have. It didn't feel like it impaired the vision or understanding of the traffic situation.'' 
    \item \textbf{Manual Drivers:} %``Als Autofahrer hatte ich das Gefühl wie mich im normalen Straßenverkehr zu befinden und zusätzlich fand ich die Signale des selbstfahrenden Autos hilfreich und bestätigend.'' / 
    ``As a driver, I felt like I was in normal road traffic and I also found the signals from the self-driving car helpful and reassuring.'' 
\end{itemize}

\paragraph{Learning and Familiarity}
There was a learning curve associated with understanding AV communications. Familiarity with the signals over time helped reduce uncertainty.
\begin{quote}
``Initially, I had to learn what the automated vehicle wanted to tell me with the message, but afterward, the interactions were easier.'' 
\end{quote}

\begin{quote}
%"Die Signale sollten bei allen autonomen Fahrzeugen die selben sein, sonst zu verwirrend und zu schwer zu lernen." / 
``The signals should be the same for all autonomous vehicles, otherwise it's too confusing and too hard to learn.''
\end{quote}

Standardizing signals across all AVs could reduce learning time and confusion.

%\subsubsection{Areas for Improvement}

\paragraph{Signal Differentiation}
There should be a clearer distinction between different types of signals.
\begin{quote}
``LEDs should be clearly different to normal lights + the turning symbol to be sure this won't be confused.'' 
\end{quote}

\paragraph{Interactive Feedback}
Incorporating more interactive and responsive signaling systems that adapt based on the road context.
\begin{quote}
%"Ich finde es angenehm, wenn man Rückmeldung erhält, ob das autonome Auto mich wirklich wahrgenommen hat und auf meine Reaktion wartet." / 
``I find it pleasant when I receive feedback on whether the autonomous car has really recognized me and is waiting for my reaction.''
\end{quote}

\section{Discussion}
In this work, we evaluated the effect of eHMIs on three different kinds of road user roles: pedestrian, manual driver, and cyclist.

\subsection{Reflections on eHMI Standard Measures}
Several measures have become de facto standards for evaluating eHMIs: duration time, trust, perceived safety, and mental demand. 

The eHMI did not reduce total duration compared to previous work~\cite{declercq2019, faas2020longitudinal, lau2021, colley_effects_2022, lanzer_interaction_2023}.
We did find increased trust, reduced mental demand, and increased perceived safety to various positive outcomes across these dependent variables, including shorter crossing times, reduced mental demand and effort, and increased perceived safety and trust, which aligns with earlier studies~\cite{declercq2019, faas2020longitudinal, lau2021, colley_effects_2022, lanzer_interaction_2023, 10.1145/3699778}. 

We did find significant differences between the road user roles. We assume that this is primarily attributable to the presence of an ``outside shield''~\cite{hollander2021taxonomy} and the speed when being a driver or cyclist compared to being a vulnerable road user as a pedestrian (e.g., see Section \ref{sec:mw_ps}).
We found that cyclists reported the lowest level of trust, significantly lower than pedestrians. While both are vulnerable road users, the dynamic nature of cycling likely contributes to this discrepancy. Cyclists operate at higher speeds than pedestrians and require longer braking distances, making the evaluation of an AV's yielding intent more time-critical and mentally demanding, as reflected in our mental demand scores. This inherent vulnerability in ``right-hook'' scenarios, combined with the technical limitations of the simulator's braking latency, likely exacerbated feelings of insecurity, resulting in lower baseline trust compared to the other groups. However, we found no significant interaction effects on these standard measures.

\subsection{Comparability of Different Road User Evaluation}
In this work, we statistically compared subjective dependent variables across three road user roles with appropriate scenarios for each of these roles. These scenarios, however, prohibited us from comparing objective dependent variables due to differences in mobility mode, distance to be traveled, and visible objects (for eye-tracking analysis). 

This generally leads to whether it is possible or appropriate to compare the effects of eHMIs (or any other intervention) across scenarios as diverse as ours, at least for subjective dependent variables. Previous work has evaluated subjective data across various scenarios, but always in the same road user role (e.g., \cite{colley2022effects}). The eHMI use case is unique here in the sense that the eHMI can or even must be appropriate to a diverse set of roles. 
The most comparable situation might be a traffic light or road markings relevant for at least manual drivers and cyclists. However, these are not designed by manufacturers but are determined by government regulations, making the possibility of exercising design freedom very limited or impossible. While it could be true in the future that standardization will be possible, bringing the eHMIs closer to traffic lights in the sense of limited design options, we assume that there will still be differences in the design as already perceivable in the vehicle turning lights (on/off vs. LED strips that turn on individual LEDs to create a pattern). 
We conclude that while this comparison has drawbacks, it is the only way to evaluate the appropriateness of eHMIs across road user roles besides analyzing each role individually and comparing the results more qualitatively.

\subsection{Diverse Road Users Do Not Need Different External Communication}
Our study was grounded in the premise that each road user type brings distinct communication demands to AV encounters. Pedestrians typically interpret intent at relatively close distances and rely heavily on explicit communication and nuanced kinematic information to judge whether it is safe to cross. In contrast, cyclists must maintain balance, anticipate vehicle trajectories while in motion, and often observe traffic from oblique viewing angles. On the other hand, manual drivers must interpret AV signals with established automotive lighting conventions, traffic rules, and expectations of vehicle behavior. These heterogeneous perceptual and cognitive demands naturally raise the question of whether role-specific eHMIs are necessary. However, introducing multiple, role-targeted eHMIs within shared traffic environments also presents substantial drawbacks. Road users often approach the same AV simultaneously, creating a scenario in which parallel, user-specific messages could introduce visual complexity, increase cognitive load, and reduce overall signal discriminability. Furthermore, role-specific systems risk fragmenting user expectations and undermining learnability, especially given that individuals frequently switch roles across short time scales (e.g., walking, cycling, and driving within a single trip). Prior work has already highlighted the scalability challenges inherent in proliferating eHMI variants~\cite{dey2020taming, 10.1145/3580585.3607167, colley2020scalability, colley2023scalability, 10.1145/3610977.3637478}, and role-specific implementations could exacerbate these issues by multiplying the number of signals a single AV would need to manage. Besides, as there is still a debate about the necessity of eHMIs~\cite{10.1145/3342197.3345320}, it was interesting to see how the participants rated the usage of eHMIs in these scenarios from the perspective of different road user roles. 

Participants rated the perceived safety and system usability significantly higher with an eHMI across all scenarios. Therefore, the same eHMI can be transferred to different road users, signaling the same message. This eHMI design was already conceptualized by \citet{dey2018interface} and the transferability was proposed by \citet{10.1145/3313831.3376884}. Although they only conceptualized and proposed this for AV-Cyclist and AV-Pedestrian interaction, our study shows that the interaction between an AV and a driver is also positively influenced by the same eHMI. However, unlike \citet{10.1145/3313831.3376884}, in our study, every scenario between the road user and an AV communicated a yielding message. This certainly has to be explored with other messages, such as messages that intend not to yield to a pedestrian/cyclist/manual driver. The data also corroborates prior research that eHMIs are a relevant addition to future AV communication, especially in critical situations where the right-of-way is ambiguous or unclear. 

\subsection{Practical Implications}
In our study, we isolated one focal interaction per scenario to examine role-dependent responses with greater experimental control. However, AVs will routinely operate in complex, multi-party environments where yielding or non-yielding intentions must be legible to heterogeneous users at once. The rationale for studying a unified eHMI is precisely rooted in this real-world challenge: if an eHMI is to function meaningfully in mixed traffic, it must remain interpretable across pedestrians, cyclists, and drivers without role-specific tailoring. Our findings suggest that this is feasible --- a unified eHMI design shows promise to be effectively applied across various types of road users, thereby supporting the development of a more `holistic' approach to eHMIs --- one that offers a unified solution towards diverse interactions for diverse road users~\cite{Dey2024multimodal}. This can lead to a streamlined design process, enabling AV manufacturers and designers to focus on developing a standardized interface. This could lead to faster adoption of eHMIs as there would be less need for customization and testing across different user groups. Instead, more diverse situations could be tested~\cite{colley2023scalability, 10.1145/3580585.3607167}. Furthermore, these findings could encourage regulators to adopt universal standards for eHMIs (see first efforts by the ISO~\cite{iso23049}). With clear evidence that a unified eHMI can work across various interactions, there is a stronger case for creating industry-wide guidelines, which would, in turn, accelerate the integration of eHMIs into AVs globally.

This further points to potential long-term benefits such as reduced development costs: by creating a `one-size-fits-all' eHMI, manufacturers can reduce R\&D and production costs associated with developing multiple interfaces for different road users. From a user experience point of view, such a `holistic eHMI' approach would promote consistent communication in mixed traffic situations with multiple types of traffic participants. Furthermore, road users would only need to learn one set of signals to understand interactions with AVs, regardless of whether they are driving, walking, or cycling (similar to road markings and traffic lights) --- leading to a uniform learning curve. This could lead to quicker adaptation and a smoother transition period as AVs become more common.

% No auditory signals were so far included - missing accessibility for people with vision impairments - 

These implications highlight the potential for our findings to not only influence the design and implementation of eHMIs but also to have broader impacts on the adoption of AVs and the future of urban mobility.

Furthermore, the timing of the signal proved as critical as the signal design itself. Participant feedback indicated that eHMIs triggered only upon coming to a full stop were perceived as ``late,'' merely confirming the vehicle's kinematics rather than aiding the crossing decision. Such latency can degrade trust and even mislead users who expect earlier communication. Consequently, we recommend that future eHMI standards incorporate ``pre-signals'' that trigger during the approach phase (e.g., based on a Time-to-Arrival of 3--5 seconds) rather than at a standstill. This earlier activation window is essential to allow road users sufficient time to process the message and initiate action before the vehicle enters the critical conflict zone.

\subsection{Limitations}
The study primarily involved participants aged between 21 and 30, most of whom were students or had academic experience, limiting its demographic scope. Having predominantly German participants means that cultural differences were not explored, impacting the generalizability of the results to other cultural contexts, as eHMI effectiveness can vary significantly across different regions. Future work should re-evaluate these scenarios with a more diverse sample.

In terms of transferability to real-world applications, despite efforts to create a realistic VR environment—complete with urban scenery, ambient sounds, and virtual interactions—the experience might not fully replicate actual conditions (see \citet{10.1145/3340555.3353741}). One participant's need to withdraw due to discomfort highlights this gap.

The bicycle simulator's realism was critiqued, particularly the braking system, which was perceived as too slow, potentially affecting safety perceptions. Only one type of eHMI was tested, limiting the evaluation of various interfaces that could affect user interaction with non-yielding versus yielding vehicles.

Additionally, while participants suggested colors like amber or red, these carry specific traffic semantics (e.g., turn signals, stop) that must be avoided to prevent confusion. Future work could test diverse colors and symbols, accounting for cultural differences in color interpretation and the visual contrast between the eHMI and the vehicle body (Vehicle Color $\times$ eHMI Color).

Furthermore, the study's design included a limited number of scenarios to maintain manageability and comparability, with each road user experiencing only one specific scenario. This restricted approach might not capture the full potential and challenges of different eHMI applications. 

A key limitation of our study is the short duration of each VR exposure (around 20–30\,s per scenario). While this design choice helped to keep the total session length near 75 minutes and to limit fatigue and simulator sickness, it also constrains the extent to which our findings generalize to longer, more naturalistic interactions. Future work should examine extended exposure and repeated sessions to assess adaptation, learning effects, and potential changes in perceived demand and trust over time.

An additional limitation concerns the cultural and contextual specificity of yielding behavior. Right-of-way norms, driver–pedestrian negotiation practices, and expectations about vehicle yielding vary substantially across countries and traffic cultures~\cite{10.1007/978-3-030-22666-4_37, lanzer2020designing, 10.1145/3699778}. Because our study was conducted within a German traffic context, the interpretation of yielding cues, both kinematic and eHMI-based, may reflect local conventions. Future work should investigate how cultural differences in road-user expectations and yielding norms influence the perception, learnability, and effectiveness of eHMIs in ambiguous interactions.

Lastly, the learnability aspect of the scenarios presented another limitation. Participants quickly became familiar with the AV's behavior, particularly its consistent stopping points, which could artificially enhance performance over repeated trials. The study attempted to mitigate these learning effects through a Latin Square design, but the repetitive nature of scenarios for each road user type still likely influenced the results.

%A paragraph on future work?

\section{Conclusion}
The insights from our VR-based study evaluating the subjective and objective impact of an eHMI communicating an AV's intention to different kinds of road users (pedestrians, cyclists, and drivers) under various levels of distraction shed light on the effectiveness of eHMIs in facilitating communication between AVs and diverse road users. Our findings indicate that a unified eHMI design can be successfully applied across different types of road users, enhancing perceived safety, trust in AVs, and system usability and reducing mental demand across diverse interaction scenarios, including distractions and interference. The absence of significant differences in distraction levels and the importance of the road user underscores the robustness of communication through eHMIs and points to the promise of a holistic approach to designing a unified eHMI to cater to different road users. However, the potential for confusion in scenarios involving multiple road users highlights the need for further research, particularly in interactions between AVs and manual drivers or cyclists.

\section*{Open Science}
All data and evaluation scripts are publicly available via \url{https://github.com/M-Colley/ehmi-for-all-chi26-data}.

\begin{acks}
The authors thank all study participants.
\end{acks}

\appendix

\section{Introductory Texts}\label{app:intro_texts}

This is the text for the pedestrians:

\begin{quote}
\textit{You are standing on a two-lane road. This road has mixed traffic, meaning there are both manually operated and highly automated vehicles on the road. One of the vehicles will stop to let you cross the street. Depending on the situation, this vehicle will communicate with you differently. Your goal is the green-marked zone.}
\end{quote}

This is the text for the drivers:

\begin{quote}
\textit{You are sitting in a manually driven car. This road has mixed traffic, meaning there are both manually operated and highly automated vehicles on the road. Your goal is to turn left at the intersection. Your destination is the green-marked zone.}
\end{quote}

This is the text for the cyclists:

\begin{quote}
\textit{You are a cyclist standing on a two-lane road. This road has mixed traffic, meaning there are both manually operated and highly automated vehicles on the road. You want to go straight across the intersection. Your goal is the green-marked zone.}
\end{quote}

This is the text for every scenario with an eHMI:

\begin{quote}
\textit{The automated vehicles will communicate with you through an additional display. The LED strip will blink as soon as the vehicle intends to let you cross.}
\end{quote}

This is the text for every scenario without an eHMI:

\begin{quote}
\textit{The automated vehicles will not communicate with you through an additional display.}
\end{quote}

\section{Check for Order Effects}\label{app:order}

\begin{figure*}[ht]
    \centering
    \includegraphics[width=1.00\linewidth]{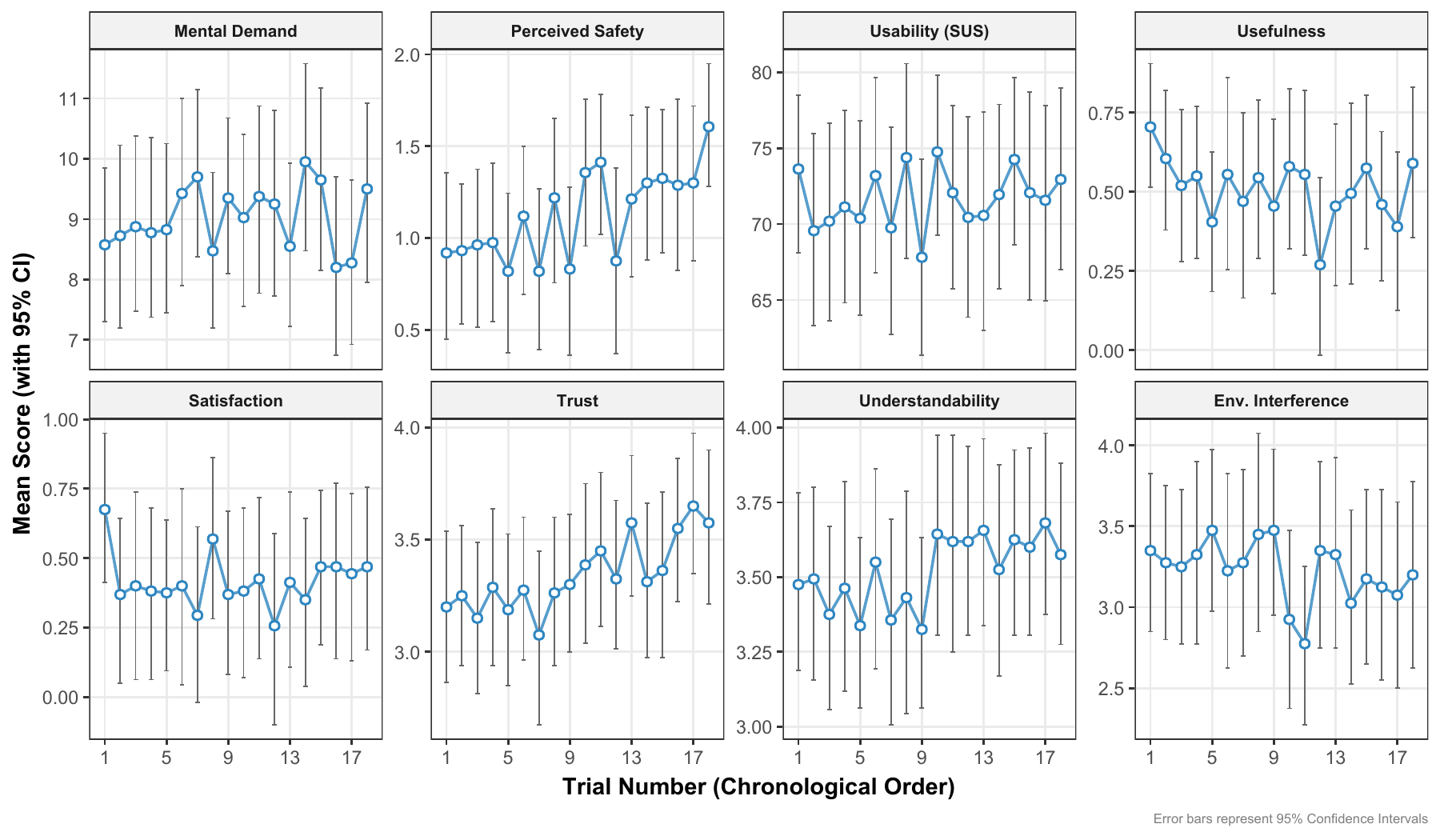}
    \caption{Investigation of order effects: Mean scores plotted against trial number (1--18) for all subjective measures. Error bars indicate 95\% confidence intervals. No significant correlations or trends were found, confirming that the counterbalancing was effective and results were not confounded by participant fatigue or learning.}
    \label{fig:order_effects_by_trial_clean}
    \Description{Investigation of order effects: Mean scores plotted against trial number (1--18) for all subjective measures. Error bars indicate 95\% confidence intervals. No significant correlations or trends were found, confirming that the counterbalancing was effective and results were not confounded by participant fatigue or learning.}
\end{figure*}

\bibliographystyle{ACM-Reference-Format}
\bibliography{sample_1.bib}

\end{document}